\documentclass[english]{elsart}
\usepackage[latin2]{inputenc}
\usepackage[T1]{fontenc}
\usepackage[english]{babel}
\usepackage{amsmath}
\usepackage{setspace}
\usepackage{amssymb}
\usepackage{color}
\makeatletter
\providecommand{\LyX}{L\kern-.1667em\lower.25em\hbox{Y}\kern-.125emX\@}
\DeclareMathOperator{\distr}{distr}
\DeclareMathOperator{\diag}{diag}

\usepackage{setspace}
\usepackage[numbers]{natbib}
\usepackage[dvips]{graphicx}
\usepackage{epic,epsfig}
\usepackage{subfigure,float}
\usepackage{amsmath,amsfonts,amssymb}
\usepackage{inputenc}
\usepackage{natbib}
\usepackage{euscript}
\makeatother

\begin{document}
\begin{frontmatter}

\title{Spectral methods and cluster structure in correlation-based networks}

\author{Tapio Heimo$^{1}$},
\author{Gergely Tibély$^{2}$\corauthref{cor1}},
\ead{tibelyg@maxwell.phy.bme.hu}
\author{Jari Saram\"aki$^{1}$},
\author{Kimmo Kaski$^{1}$}, and
\author{János Kertész$^{1,2}$}
\corauth[cor1]{Corresponding author.}

\address{$^{1}$ Laboratory of Computational Engineering, Helsinki
  University of Technology, P.O. Box 9203, FIN-02015 HUT, Finland}

\address{$^{2}$ Department of Theoretical Physics, Budapest University
of Technology and Economics, Budafoki ut 8, H-1111 Budapest, Hungary}

\begin{abstract}
  
We investigate how in complex systems the eigenpairs of the matrices
derived from the correlations of multichannel observations reflect the
cluster structure of the underlying networks. For this we use daily
return data from the NYSE and focus specifically on the spectral
properties of weight $W_{ij} = |C|_{ij} - \delta_{ij}$ and diffusion
matrices $D_{ij} = W_{ij}/s_j- \delta_{ij}$, where $C_{ij}$ is the
correlation matrix and $s_i = \sum_j W_{ij}$ the strength of node
$j$. The eigenvalues (and corresponding eigenvectors) of the weight
matrix are ranked in descending order. In accord with the earlier observations the first
eigenvector stands for a measure of the market correlations. Its
components are to first approximation equal to the strengths of the
nodes and there is a second order, roughly linear, correction. The
high ranking eigenvectors, excluding the highest ranking one, are
usually assigned to market sectors and industrial branches. Our study
shows that both for weight and diffusion matrices the eigenpair
analysis is not capable of easily deducing the cluster structure of
the network without \textit{a priori} knowledge. In addition we have studied
the clustering of stocks using the asset graph approach with and
without spectrum based noise filtering. It turns out that asset graphs
are quite insensitive to noise and there is no sharp percolation
transition as a function of the ratio of bonds included, thus no
natural threshold value for that ratio seems to exist. We suggest that
these observations can be of use for other correlation based networks
as well.

\end{abstract}


\end{frontmatter}

\section{Introduction}

The network approach to complex systems has turned out to be extremely
fruitful in revealing their structure and function
\cite{Newman_Barabasi_Watts,Dorogovtsev_Mendez,NewmanReview,CaldarelliBook}.
The usual way to construct the network is to identify
the elements of the system with nodes, between which the links are present if 
the corresponding interactions exist. In the case of weighted networks, the 
weight of a link is identified with the strength of the interaction. 

Processes taking place in a complex system, represented as a network, depend
heavily on its structure. For example motifs that are statistically 
significantly overrepresented as compared to a random reference system are 
supposed to have some functional role \cite{UriAlon,OnnelaIntensity}. 
Moreover, communities i.e. groups that are well wired internally but loosely 
connected to the rest of the network, play an eminent role in dynamic
phenomena like spreading \cite{newman2,CliquePerc,WeakLinks}. Clearly, the
investigation of the network structure is of central interest.

For many systems, however, the nature of interactions is
hidden and only some activities of the nodes can be measured, e.g., in the form
of time series. For such systems, the natural network representation
is a complete graph with weights corresponding to the elements of the 
correlation matrix determined by the nodal activities. Then the task is to 
filter out from the noisy correlation matrix the groups of closely related 
elements. This problem is quite general and it appears in many fields of 
research ranging from the evaluation of micro-array data to portfolio 
optimization. In this paper we have chosen to study correlation matrices of 
stock returns, but we think that the network approach and the observations 
made here have also more general validity.

Correlations between time-series of stock returns serve as one of the main inputs
in the portfolio optimization theory. 
In the classical Markowitz portfolio optimization the correlations are
used as measures of the dependence between the time series and the variance as the measure of risk
\cite{markowitz}.\footnote{There are conceptual problems with this
approach, which we do not want to address here \cite{bouchaud}.}
As the empirical time series are always finite, the resulting
correlation matrix is noisy. This brings up the need to reduce the
noise, for which the most frequently applied tool in the
financial literature is principal component analysis \cite{Tsay}. 

Previously, correlation matrices of stock return time series have
been studied from the network point of view, e.g., by using maximal 
spanning trees (MST). The maximal spanning tree of a network is a tree
containing all the $N$ nodes and $N-1$ links such that the sum of the
weights is maximized. It was introduced in the study of financial
correlation matrices by Mantegna \cite{mantegna1}, who was able to 
identify groups of stocks that make sense from an economic point of view. It was discovered 
that often, the branches of the MST correspond to business sectors or
industries. Moreover, this method enables to describe the hierarchical
organization of the market  and has been applied to monitor 
the effect of the time dependence of the correlations \cite{onnela:dyn}
Later, MSTs of diverse financial correlation based networks have been studied, e.g., 
by Bonanno \emph{et al.} \cite{bonanno1, bonanno2, bonanno3} and 
Onnela \emph{et al.} \cite{onnela:dyn, onnela2}. 
Indeed, the MST method is simple and gives reasonable results. However, it is too 
restrictive and thus other, complementary methods are needed.

In the so called asset graph approach, one ranks the links according to the
values of the corresponding correlation matrix and considers only a
fraction $p$ of the strongest ones as occupied. By using this method for low 
values of $p$ Onnela \emph{et al.} \cite{onnela:clust} and Heimo
\emph{et al.} \cite{Tapio_Tokyo} found clear evidence of strong
intra-business sector clustering. It has also been suggested that planar 
maximally filtered graphs yield a natural extension to the MST 
approach \cite{tumminello}. Other interesting approaches include methods 
based on the super-paramagnetic Potts model \cite{SPpotts} and on the 
maximum likelihood optimization \cite{MarsiliML}. Several financial markets
have been studied from the above points of view.

Based on these approaches a following picture about the organization
of the correlation network of the stocks  emerges: i) There is a dominant
correlation among most of the stocks reflecting the overall
behavior of the market (this is the basis of the one factor model
\cite{bouchaud}); ii) The stocks are organized hierarchically in
clusters, which mainly correspond to industrial branches (as assumed
in the multi-factor models) \cite{bouchaud} iii) There are systematic deviations
from this oversimplified picture, partly because of the ambiguous
nature of any classification scheme and partly because of inter-cluster
relations; iv) In spite of considerable robustness in the
correlations during the time of "business as usual", major events like
crashes cause dramatic changes in the network structure \cite{onnela:dyn}.

All information on the network structure is encoded in its adjacency
matrix, or, for weighted networks, in the weight matrix. Likewise, all
information on the structure of correlations of stock returns is to be
found in the correlation matrix. Consequently, this information is
also inherited by the eigenvalues and eigenvectors of such
matrices. If the data is structured in terms of clusters and communities
that these matrices represent, it should also be reflected in the
eigenpairs.  For financial correlation matrices, it has been shown that 
most eigenpairs correspond to noise accessible by random matrix theory and thus the information about 
the cluster structure is contained in a few non-random eigenpairs
\cite{RMT_Potters,RMT_Stanley,Stanley_long} (see \cite{Burda_rev} for
an overview).  It was also suggested that clusters of highly correlated
stocks could be identified by studying 
the localization of non-random eigenvectors. In this paper we
investigate the questions of how the eigenpairs of correlation and
related matrices 
derived from stock price time series reflect the cluster structure and 
industry sectors.

This paper is organized as follows. Section \ref{sec:basics} gives a
short introduction to the spectral properties of weight and diffusion
matrices and their relationship to the cluster structure of the
underlying network. Section \ref{sec:1st} is devoted to the analysis
of the largest eigenvalue and the corresponding eigenvector, whereas
the intermediate, non-random eigenpairs are examined in Section
\ref{sec:interm}. The asset graph approach to the clustering of stocks
is discussed in Section \ref{sec:assetg}.

\section{Basic notions}
\label{sec:basics}
\subsection{Matrices related to weighted networks}

A simple undirected and weighted network can be represented by a 
\emph{weight matrix} $\boldsymbol{W}$ in which an element
$W_{ij}=W_{ji}$ ($\ge 0$ in our study) corresponds to the weight of the 
link between the nodes $i$ and $j$ and the diagonal elements are zero. Note that 
$W_{ij}=0$ signifies the absence of the link. The sum
$s_{i} = \sum_{j} W_{ij}$ is the strength of node $i$~\cite{Barrat:2004ye}.
Here, we restrict our analysis to irreducible networks, \emph{i.e.},
networks consisting of just one connected component. In this case  
the Frobenius-Perron theorem states that $\boldsymbol{W}$ has a
largest positive eigenvalue and the components of the corresponding
eigenvector are non-zero and of the same sign.\footnote{In network analysis,
these components have been interpreted as measures of centrality
of the corresponding nodes (see., \textit{e.g.},~\cite{NewmanReview,Scott}).
Here the idea is that the centrality $x_{i}$ of node $i$ should be proportional 
to the average of the centralities of its neigbours, weighted by the
weights of the connecting links. This leads to the equation
$x_{i} = \frac{1}{\lambda} \sum_{j} W_{ij} x_{j}$,
where $\lambda$ is a constant. In matrix form,
$\boldsymbol{W} \vec{x} = \lambda \vec{x}$,
and with the restriction $x_{i} \ge 0$ the only non-trivial solution
is the eigenvector corresponding to the largest eigenvalue. This
measure of centrality is often referred to as the
\emph{eigenvector centrality}. }

If the elements of $\boldsymbol{W}$ are i.i.d. random numbers with finite variance
$\sigma^{2}$, the probability density of the first eigenvalue
converges to a normal distribution \cite{RanGraphs}
\begin{equation}
\lim_{N\to\infty}\distr\left\{\lambda_1-\left[ (N-1)\mu+\sigma^2/\mu
  \right]\right\} = \mathcal{N}(0,2\sigma^2)
\end{equation}
where $\mu>0$ is the mean of the matrix elements. The first term inside the
square brackets expresses the average node strength, while
the second term is due to fluctuations. In some cases statements
can be made about the whole spectrum. We return to this in section
\ref{sec_CorMat}.

\emph{Diffusion} process in terms of random walks can be used as a tool for studying the
structure of a network \cite{NorditaPRL, Nordita, Nordita2}.  
At each time step a walker moves at random from its current node $j$ to 
node $i$ with probability $T_{ij} = W_{ij} / s_{j}$.
%The diffusion process describes the dynamics of random walkers moving on
%the network such that at each time step a walker moves from its
%current node $j$ to node $i$ with probability $T_{ij} = W_{ij} / s_{j}$.
If we denote the walker density at node $i$ by $v_{i}(t)$, the
average dynamics of the process is described by
\begin{equation}
\label{eq:transfer}
\vec{v}(t+1) = \boldsymbol{T}\vec{v}(t)
\end{equation}
or equivalently by
\begin{equation}
\label{eq:diffusion}
\vec{v}(t+1) - \vec{v}(t) = \boldsymbol{D} \vec{v}(t),
\end{equation}
where $\vec{v}(t) = (v_{1}(t), \ldots,v_{N}(t))^{T}$ and $\boldsymbol{D} =
\boldsymbol{T} - \boldsymbol{I}$. Here, $\boldsymbol{T}$ denotes the \emph{transfer
  matrix} and $\boldsymbol{D}$ the \emph{diffusion matrix}. These matrices
have clearly the same eigenvectors and the spectrum of $\boldsymbol{D}$
is identical with the spectrum of $\boldsymbol{T}$ shifted to the left by
unity. The matrix $\boldsymbol{T}$ can be mapped into a symmetric matrix by 
the similarity transformation 
$\diag(s_i^{-1/2})\cdot\boldsymbol{T}\cdot\diag(s_i^{1/2})$ and
therefore its eigenvalues are real. Futhermore, the walker density cannot diverge 
at any node, so the eigenvalues of $\boldsymbol{T}$  must lie within the 
interval $[-1,1]$. The strength vector $\vec{s} = (s_{1}, \ldots, s_{N})$ is an 
eigenvector of $\boldsymbol{T}$ with eigenvalue $1$ and due to the Frobenius-Perron 
theorem this eigenvalue is non-degenerate. Thus
\begin{equation}
\label{eq:stationary}
\lim_{t \to \infty} \vec{v}(t) =  \vec{s}, 
\end{equation}
unless the network is bipartite, in which case $-1$ is also an eigenvalue. 

In addition to walker densities $v_{i}(t)$, the diffusion process can
also be analyzed by studying the walker densities per
unit strength defined by
\begin{equation}
c_{i}(t) = \frac{v_{i}(t)}{s_{i}}.
\end{equation}
It is straightforward to show that
\begin{equation}
\vec{c}(t+1) = \boldsymbol{N} \vec{c}(t),
\end{equation}
where $\boldsymbol{N} = \boldsymbol{T}^{T}$ is called the \emph{normal
  matrix}. Clearly the only difference
between the governing equations for the densities and the densities
per unit strength is that $\boldsymbol{T}$ is replaced with $\boldsymbol{N}$. 
From Eq. (\ref{eq:stationary}) we see that
\begin{equation}
\label{eq:stationary2}
\lim_{t \to \infty} \vec{c}(t) =  (1, \ldots, 1)^{T}. 
\end{equation}

\subsection{Modular structure and eigenvectors} \label{sec:eigvec}

Recently there has been increasing interest in the "mesoscopic" properties
of networks, \emph{i.e.}, in structures beyond the scale of single vertices or 
their immediate neighborhoods. One important related problem is the detection
and characterization of \emph{modules} or \emph{communities}
 \cite{newman2,CliquePerc,Guimera,Reichardt,Fortunato,Jussi}, which
are, loosely speaking,
groups of vertices with dense internal connections and weaker
connections to the rest of the network. 

Evidently, the weight, transfer, normal, and 
diffusion matrix representations of a modular network carry
information about the modules in their eigenvalues and -vectors.
In the case of diffusion, it is tempting to assign to the eigenvalues
and -vectors a direct physical interpretation~\cite{NorditaPRL,
  Nordita, Nordita2}. If a random walker enters a
module with dense internal connections and sparse connections to the rest
of the network,  it  gets, on average, ``trapped'' for a long time. This
phenomenon is reflected to the spectral expansion of the random
walker density at time $t$,
\begin{equation} \label{eq:rw}
        v_i(t)=\sum_j c_j \lambda_j^t\cdot
        [\vec{e}_j]_i
\end{equation}
\begin{equation}  \label{eq:c}
        \vec{c}=\boldsymbol{E}^{-1} \vec{v}(0),
\end{equation}
where $\boldsymbol{E}$ contains the eigenvectors $\vec{e}_{j}$ of the 
transfer matrix in its columns.
If convergence to the stationary state is slow,
Eq. (\ref{eq:rw}) should contain some terms with eigenvalues close to
$1$ or $-1$. Eigenvalues close to $-1$ indicate that the network is
almost bipartite. On the other hand, large positive eigenvalues are
consequences of modular structure and the corresponding
eigenvectors can be expected to carry information about the structure
of communities. 

The interpretation of the eigenpairs of the weight matrix is more
difficult. However, one can naturally iterate a vector $\vec{v}$ 
(a ''phantom field'' on the nodes of the network) by the weight matrix, 
and study its properties. Eq. (\ref{eq:rw}) still applies, and 
Eq. (\ref{eq:c}) can be written in a simpler form $c_j=\vec{e}_j\cdot\vec{v}(0)$ 
due to the symmetry of the weight matrix (of course, $\vec{e}_{j}$ are now
eigenvectors of the weight matrix). Here a convenient
initial condition is $v_{j}(0)=\delta_{ij}$, and as $v_a(t+1)=\sum_jW_{aj}v_j(t)$,
the new value of this quantity on node $a$ will be a weighted sum of
the (old) values on $a$'s neighbours. If the spreading of this quantity starts from
node $i$, located in a densely interconnected module, during the
first time steps only the other members of this module get significant
contribution, as
$i$ and its neighbours have most of their links within the
module. The phenomenon resembles the ``trap''-behaviour of the
modules in the case of diffusion. Noticing that\footnote{Here, we must
assume that $\vec{v}(0)$ is not perpendicular to $\vec{e}_1$.}
\begin{equation}  \label{eq:conv}
  \frac{\vec{v}(t)}{\lambda_1^t} =
  \frac{\boldsymbol{W}^t\vec{v}(0)}{\lambda_1^t}
  \xrightarrow[t\to\infty]{} \vec{e}_1,
\end{equation}
where $\lambda_{1}$ is the largest eigenvalue of $\boldsymbol{W}$,
we see that  the ratios $v_i/v_j$ approach to constants. The speed of
the convergence depends naturally on the magnitudes 
of the other eigenvalues. 

The fact that eigenvalues close to the largest one
slow down the convergence, suggests that the corresponding eigenvectors
can carry information about the modules, similarly to the
eigenvectors of the diffusion matrix. In conclusion, it seems to be
reasonable to interpret the eigenvectors of the weight matrix
similarly to the eigenvectors of the diffusion matrix, at least from
the point of view of network modularity.

Now, let us turn shortly to the interpretation of eigenvector components, both 
for diffusion and weight matrices. Consider a case in which
the eigenvectors are ranked in descending order and $\lambda_2 \gg
|\lambda_{3\ldots N}|\approx0$. Assume that the second
eigenvector is localized on two sets of nodes, such that the
eigenvector components corresponding to the first set are positive,
and the components corresponding to the second set are negative. Then,
the second term in Eq. (\ref{eq:rw}) gives a slowly decaying correction
for both sets of nodes, but with different signs. This means that the
random walker or the ``phantom field'' gets trapped for a while in
one set, and is held back from entering into the other set. So the
first set of nodes can be thought of as  a community. Changing the
initial condition such that $c_2$ changes its sign, and applying the 
above arguments shows that the other set can also be thought of as a 
community. Both of these cases show that the two communities are far from 
each other as regards to the average travelling time of a random walker 
between them.  It should be noted here that using absolute values or 
squares of the eigenvector components (e.g. \cite{Stanley_sectors}, 
\cite{Jeong}) is clearly inappropriate, as an eigenvector may be 
localized on two extremely distant communities.

As mentioned in the previous section the diffusion process can also be
analyzed by studying the time evolution of the walker densities per unit strength
$c_{i}(t)$. Simonsen \emph{et al.} have suggested that the
eigenvectors of the normal matrix $\boldsymbol{N}$ corresponding to the largest
eigenvalues contain a lot of information about the modular structure of the 
network and that the modules can be identified with the so called current mapping
technique \cite{NorditaPRL, Nordita, Nordita2}. Here, one should
notice that the $i$th component of the $k$th eigenvector of
$\boldsymbol{N}$ is equal to $[\vec{e}_{k}]_{i}/s_{i}$, where $[\vec{e}_{k}]_{i}$
is the $i$th component of the $k$th eigenvector of the transfer matrix $\boldsymbol{T}$.

\subsection{Correlation Matrices}
\label{sec_CorMat}

The equal time correlation matrix $\boldsymbol{C}$ of $N$ variables
can be estimated from $T$ observations by
\begin{equation} \label{eq:corr}
C_{ij}=\frac{\langle \mathbf{r}_{i} \mathbf{r}_{j}
  \rangle -\langle \mathbf{r}_{i} \rangle \langle \mathbf{r}_{j} \rangle }
  {\sqrt{[\langle {\mathbf{r}_{i}}^{2} \rangle
      -\langle \mathbf{r}_{i}\rangle ^{2}][\langle {\mathbf{r}_{j}}^{2} 
       \rangle -\langle \mathbf{r}_{j} \rangle
      ^{2}]}}, 
\end{equation}
where $\mathbf{r_i}$ is a vector containing the observations of the
variable $i$. In the case of Gaussian i.i.d. variables,
$\boldsymbol{C}$ is the Wishart matrix and its eigenvalue density
converges as $N\to\infty$, $T\to\infty$, while $N/T \le 1$ is fixed, to
\cite{MarchPast}
\begin{equation} \label{eq:MP}
  \rho_{\boldsymbol{W}}(\lambda)=\left\{ \begin{array}{ll}  
    \frac{T/N}{2\pi\sigma^2}\frac{\sqrt{(\lambda_{\textrm{max}}-\lambda)(\lambda-\lambda_{\textrm{min}})}}
       {\lambda} &  \textrm{if }\lambda_{\textrm{min}}\leq\lambda\leq\lambda_{\textrm{max}}\\
    0 & \textrm{else} \end{array}\right.
\end{equation}
\begin{equation}        \label{eq:MP2}
  \lambda_{\textrm{max/min}}=\sigma^2\left(1\pm\sqrt{N/T}\right)^2,
\end{equation}
where $\sigma^{2} = 1$ due to the ``normalization'' in Eq. (\ref{eq:corr})\footnote{Without the normalization, $\sigma^{2}$ would be the variance
of the variables.}.
In empirical cases, significant deviations from Eq. (\ref{eq:MP}) can
usually be considered as signs of relevant information \cite{Stanley_long}.

A correlation matrix can be transformed to \emph{weight matrix} of a simple
undirected and weighted network by
\begin{equation} \label{eq:CW}
W_{ij} = |C_{ij}| - \delta_{ij}.
\end{equation}
From the point of view of network theory, the transformation can be
justified by interpreting the absolute values as measures of interaction 
strength without considering whether the interaction is positive or negative.
If the elements of $\boldsymbol{C}$ are non-negative,
$\boldsymbol{W}=\boldsymbol{C}-\boldsymbol{I}$. Therefore, the
transformation does not change the eigenvectors but the eigenvalues
are shifted to the left by unity. The correlation matrices studied in this
paper, however,  contain a few slightly non-negative
elements. Fortunately, taking the absolute
value of the elements does not change the numerical values of the
spectral quantities significantly.

In the following sections, we study correlation matrices constructed
from the logarithmic returns of New York Stock Exchange (NYSE) traded
stocks. We use two different data sets. The larger one consists of the
daily closing prices of 476 stocks and ranges from 2-Jan-1980 to 31-Dec-1999. In
the smaller data set the number of stocks is 116 and the time window
ranges from 13-Jan-1997 to 29-Jan-2000.
With both data sets the length of the time series $T$ is not very
large compared to the number of stocks $N$. Therefore the
correlation matrices are noisy.

\section{First eigenpair}        \label{sec:1st}

The largest eigenvalue of a correlation matrix derived from stock
return time series is always clearly separated from the rest of the
spectrum. The corresponding eigenvector is typically interpreted to be 
representative of  the whole market \cite{Stanley_long}, and is usually
called the \emph{market eigenvector}.

\subsection{Approximations of the first eigenvector}  \label{sec:appr}

Perhaps the simplest way to approximate the first eigenvector of the weight
matrix is to iterate a vector that is not perpendicular to the first
eigenvector by the weight matrix (see Eq. (\ref{eq:conv})). From the
Frobenius-Perron theorem we know that the components of the first
eigenvector have the same sign, so a natural choice for the initial vector is
one with uniform components. The first iteration of this vector
yields a vector proportional to the strength vector $\vec{s}$. Since
$\vec{s}$ is the first eigenvector of the diffusion matrix derived
from the weight matrix, we see that the first
eigenvectors of the weight and the diffusion matrices are the sime after first iteration.
Of course, one can find examples where this approximation is far from the asymptotics.

Another simple way to approximate the first eigenvector of the weight
matrix  is a perturbation-based calculation. In
\cite{Jeong}, such an approach was presented, although with a
different definition of perturbation. Here we separate the empirical weight matrix
into two terms as
\begin{equation}
 W_{ij}=(1-\delta_{ij})w_0+V_{ij}, 
\end{equation} 
where $w_0$ is the average off-diagonal element of the weight
matrix and $V_{ij}$ is the deviation considered as perturbation. 
The eigenvalues of the unperturbed matrix are
\begin{equation}   \label{eq:l1}
\lambda_1^{(0)}=(N-1)w_0,
\end{equation}     
\begin{equation}   \label{eq:l_others}
\lambda_{2\ldots N}^{(0)}=-w_0,
\end{equation}
and the first eigenvector is
\begin{equation} \label{eq:e1}
\vec{e}_1^{(0)} = \frac{1}{\sqrt{N}}(1,1,\ldots,1)^{T}
\end{equation}
The first order correction of the first eigenvalue reads
\begin{equation}         \label{eq:lambda1^(1that our)}
 \lambda_1^{(1)} = \vec{e}_1^{(0)^T}\boldsymbol{V}\vec{e}_1^{(0)} =
 \frac{1}{N} \sum_{ij}{V_{ij}}=0,
\end{equation}
and thus $\lambda_{1} \approx (N-1) w_{0}$, assuming that the
perturbation expansion converges fast. The first order correction of
the first eigenvector is
\begin{equation} \label{eq:1st}
\vec{e}_1^{(1)}
=\frac{\boldsymbol{V}}{Nw_0}\cdot\vec{e}_1^{(0)}-\frac{\vec{e}_1^{(0)^T}\boldsymbol{V}\vec{e}_1^{(0)}}{Nw_0}\cdot\vec{e}_1^{(0)}
=\frac{\boldsymbol{V}}{Nw_0}\cdot\vec{e}_1^{(0)},
\end{equation}
and the $i$th component the corresponding first order approximation is
\begin{equation}    \label{eq:1stappr}
 \left[\vec{e}_1^{(0)}+\vec{e}_1^{(1)}\right]_{i} = \frac{1}{\sqrt{N}}
 (1+\frac{s_{i} - \overline{s}}{Nw_0}) \approx \frac{s_i}{\overline{s}
 \sqrt{N}},
\end{equation}
where $N \gg 1$ has been assumed.

Thus the result is similar to the one obtained by the iterative way --
in first order perturbation theory, the components of the first
eigenvector are proportional to the corresponding strengths. This
proportionality was also pointed out in Ref.~\cite{Jeong}. We add to
this observation that, provided the first order approximation is
sufficient, the average off-diagonal matrix element is proportional to
the largest eigenvalue.

\subsection{First eigenpair of financial correlation based networks}
\label{sec:first_emp}

\begin{figure}[!ht]
\begin{center}
\includegraphics*[width=180pt]{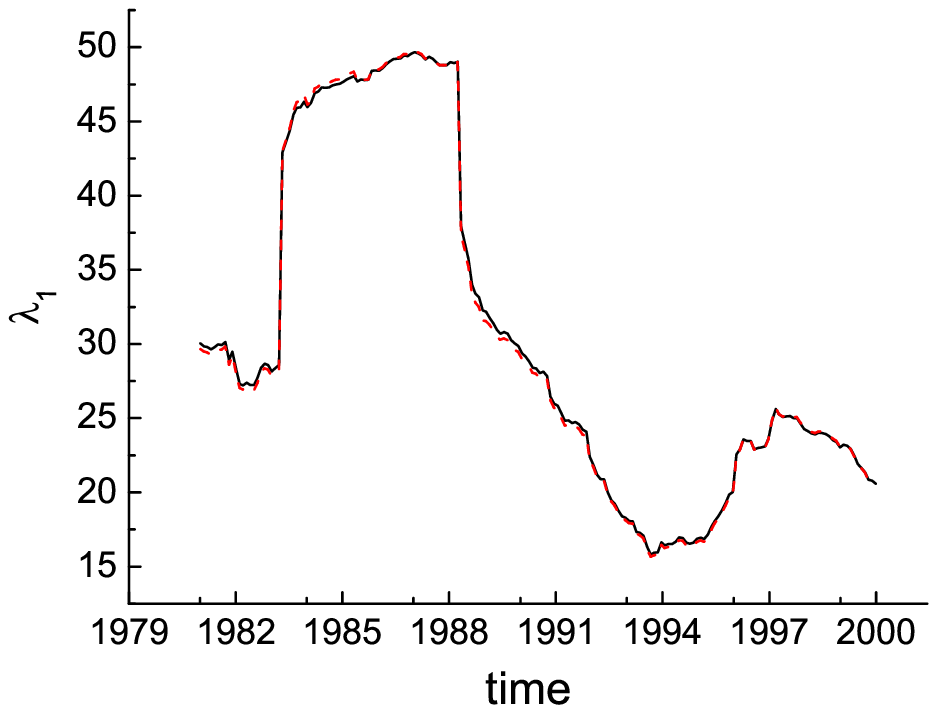}
\includegraphics*[width=180pt]{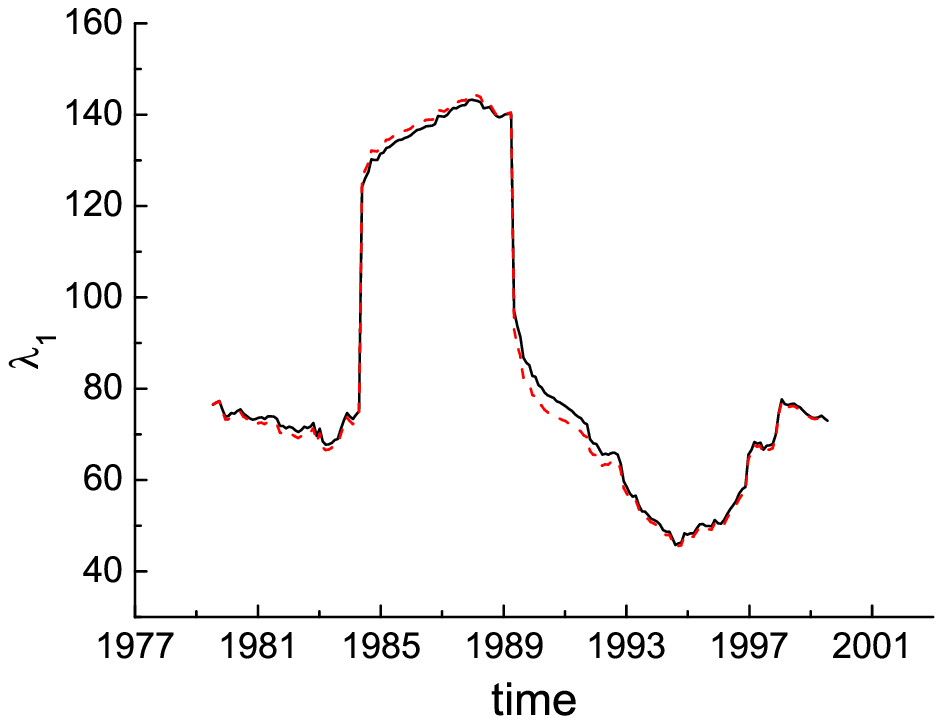}
\end{center}
\caption{(color online) The first eigenvalues (solid line) and rescaled mean
  correlations (dashed line) as functions of time for the
  $116$-stocks database (on the left) and for the $476$-stocks database 
  (on the right). The correlation matrices were constructed from 1000 trading days 
  in both cases. The outstanding plateau is a consequence of Black Monday, 
  a large market crash on October 19, 1987.}
\label{fig:l1}
\end{figure}

\begin{figure}[!h]
\begin{center}
\includegraphics*[width=180pt]{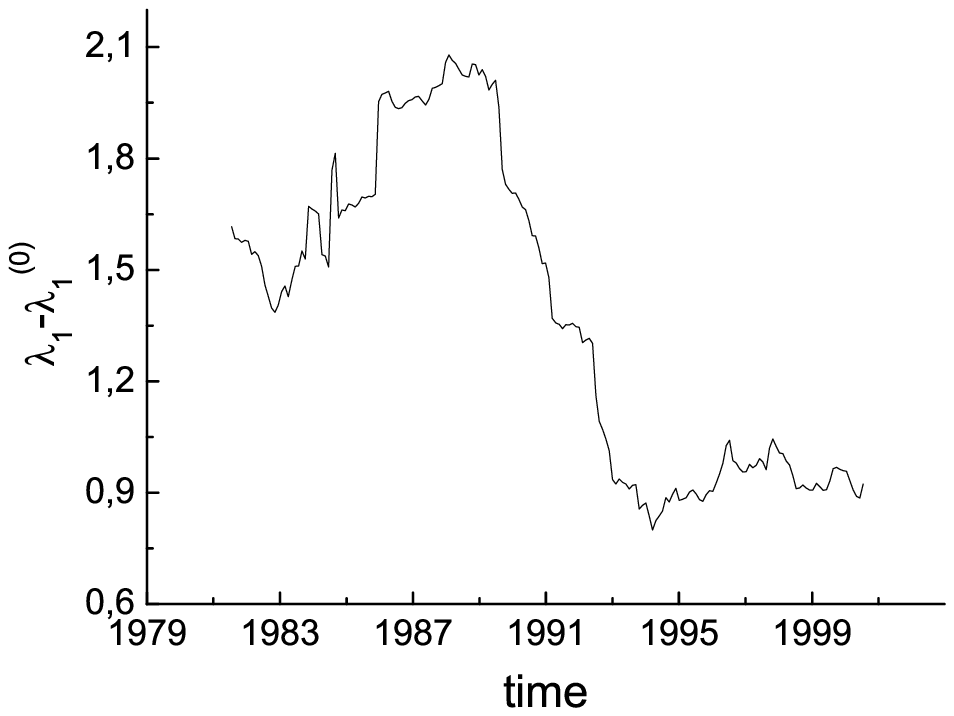}
\includegraphics*[width=180pt]{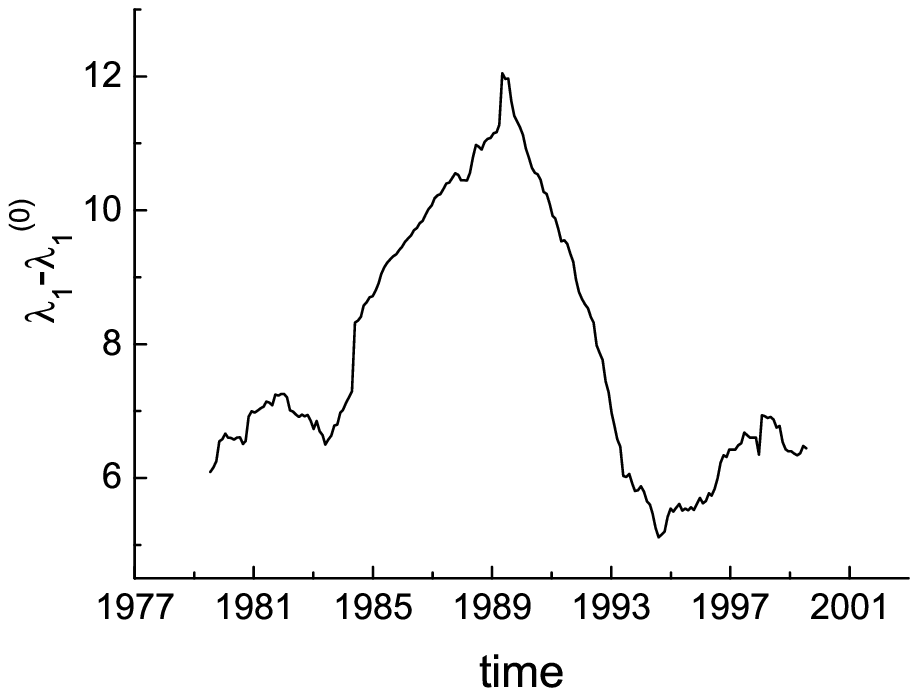}
\end{center}
\caption{Difference between the first eigenvalue and its zeroth order
  approximation for the $116$-stocks dataset (on the left) and for the
  $476$-stocks dataset (on the right). A window of 1000 trading days has
  been used.}
\label{fig:error}
\end{figure}

Based on the previous section, it is not surprising that the largest 
eigenvalue $\lambda_1$ is strongly correlated with the mean correlation 
coefficient in both data sets used. This is illustrated in Fig. \ref{fig:l1}. 
Deviations from the zeroth order approximation are illustrated in
Fig. \ref{fig:error}. 

\begin{figure}[!h]
\begin{center}
\includegraphics*[width=250pt]{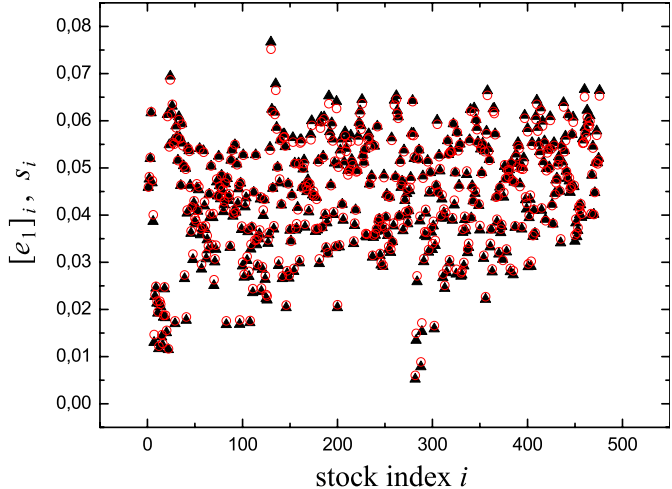}
\end{center}
\caption{(color online) Components of the first eigenvector
  ($\blacktriangle$) and the normalized strengths 
  (\textcolor{red}{$\circ$}) (the larger data set is used). The ordering of the 
stocks is such that stocks belonging to same business sector according 
to Yahoo classification \cite{yahoo} are next to each other.}
\label{fig:e1}
\end{figure}

As expected, the eigenvector corresponding to the largest eigenvalue is well
approximated by the strength vector. This is illustrated in
Fig. \ref{fig:e1}. 
\begin{figure}[!h]
\begin{center}
\includegraphics*[width=400pt]{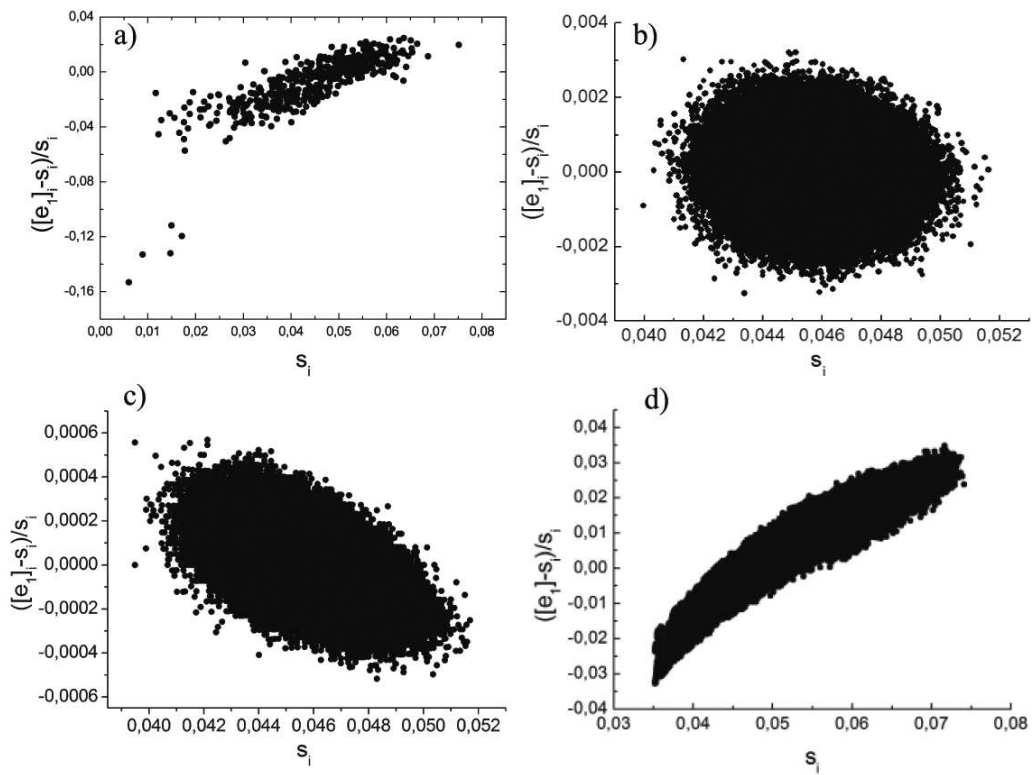}
\end{center}
\caption{The relative differences of the components of the first
  eigenvector and the (normalized) strength-vector as functions of
  the (normalized) strengths. a) The empirical data set ($N=476$
  stocks), 
b) random matrices with $i.i.d$ elements from the uniform distribution,
c) correlation matrices of the one factor model with the same length
of time series and the same mean correlation as the empirical matrix,
d) artificial multiblock correlation matrices. All results for
artificial matrices are averages over
1000 runs.}
\label{fig:reldiff}
\end{figure}
A further observation is that the \emph{relative differences} of the
components of the first eigenvector and the (normalized) strengths are
positively correlated with the strengths (Fig. \ref{fig:reldiff}, panel a). In order
to understand this effect, we have constructed and numerically
analyzed several kinds of correlation matrices. Our observations are 
as follows: the above correlation does not exist for random matrices, in which 
the elements are i.i.d.
random variables from uniform distribution
(Fig. \ref{fig:reldiff}, panel b). Surprisingly, the correlation is
negative for the one factor model with the same mean correlation, when
the correlation matrix is constructed from finite time segments of
uncorrelated Gaussian time series \cite{Pafka}
(Fig. \ref{fig:reldiff}, panel c). A strength distribution with
non-vanishing width produces similar effect. However, for multi-block
weight matrices 
with an artificial modular structure together with additional
noise,\footnote{The matrices were constructed by
  $W_{ij}=W_{ij}^0+0.1\cdot \eta_i\eta_jr$, where the communities
are represented by matrix $\boldsymbol{W}^0$ containing ten blocks of
size $45\times45$ on the diagonal, $\eta_i=|s+1|$ is a random
parameter for each node, $s$ is drawn from the
standard normal distribution, and $r$ is drawn from the standard
uniform distribution. $\boldsymbol{W}$ was normalized such that the mean
element was equal to the mean element of the empirical weight matrix.}
similar correlation is found (Fig. \ref{fig:reldiff}, panel d). 
Hence, the observed correlation could be attributed to the presence
of modular structure in the weight matrix.

There is another interesting feature in Fig.~\ref{fig:reldiff} worth
noting. The "outliers"
in the lower left corner of panel a in Fig.~\ref{fig:reldiff}
correspond to companies related to gold and silver mining, which are known to 
be extremely weakly correlated (or even negatively correlated) with the 
other participants of the market.

\section{Intermediate eigenpairs}     \label{sec:interm}

In this section we analyze the intermediate eigenvectors of the
empirical weight and diffusion matrices.\footnote{In this section, only
the larger data set is studied} We start by discussing the
problems related to defining the information carrying eigenvectors
of the weight matrix and continue by studying how the cluster structure of the network is
reflected in the localization of these eigenvectors. 
%Our main conclusion is that the cluster structure of a network is not
%inherited to the spectral quantities as cleanly as suggested in recent
%literature \cite{RMT_Potters, RMT_Stanley, Stanley_long}. 
Lastly, we analyze the intermediate eigenvectors of the
diffusion matrix and find the highest ranking ones to be
very close to those of the weigth matrix. We also demonstrate the use
of the current mapping technique with our data set.

\subsection{Defining the intermediate eigenpairs of the weight matrix}

The highest ranking eigenpairs of the correlation matrix constructed from stock
return time series are far from being random \cite{RMT_Potters, RMT_Stanley}, but the randomness
increases rapidly together with increasing rank (on average) \cite{Tapio_Tokyo}. 
Therefore, there is no strict
border between the random and intermediate parts of the spectrum and
the identification of the information carrying eigenvectors is a
highly non-trivial task.

\begin{figure}[!h]
\begin{center}
\includegraphics*[width=180pt]{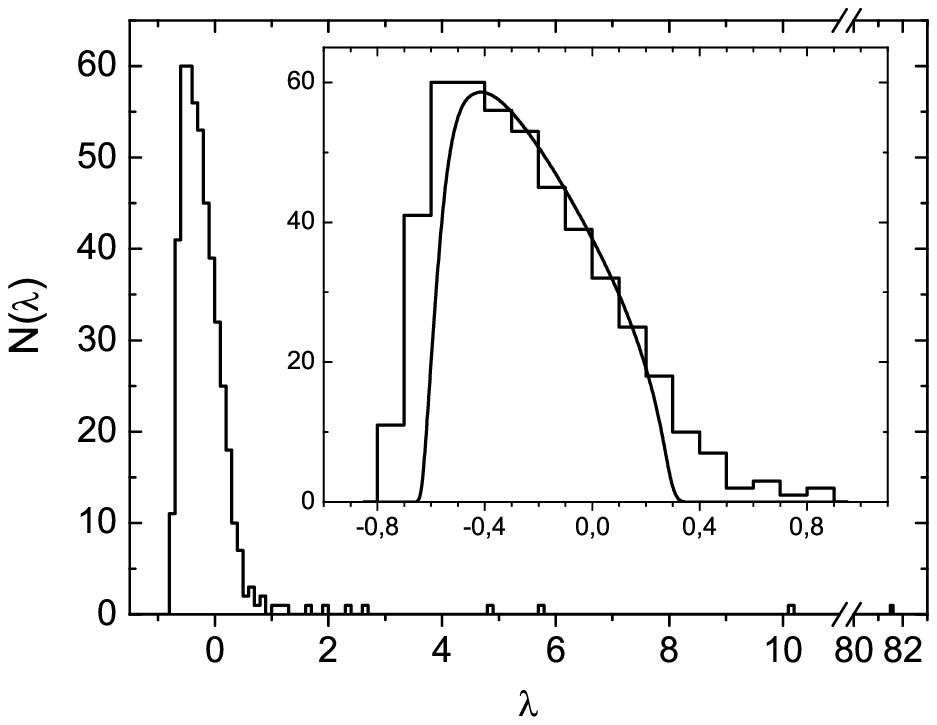}
\includegraphics*[width=180pt]{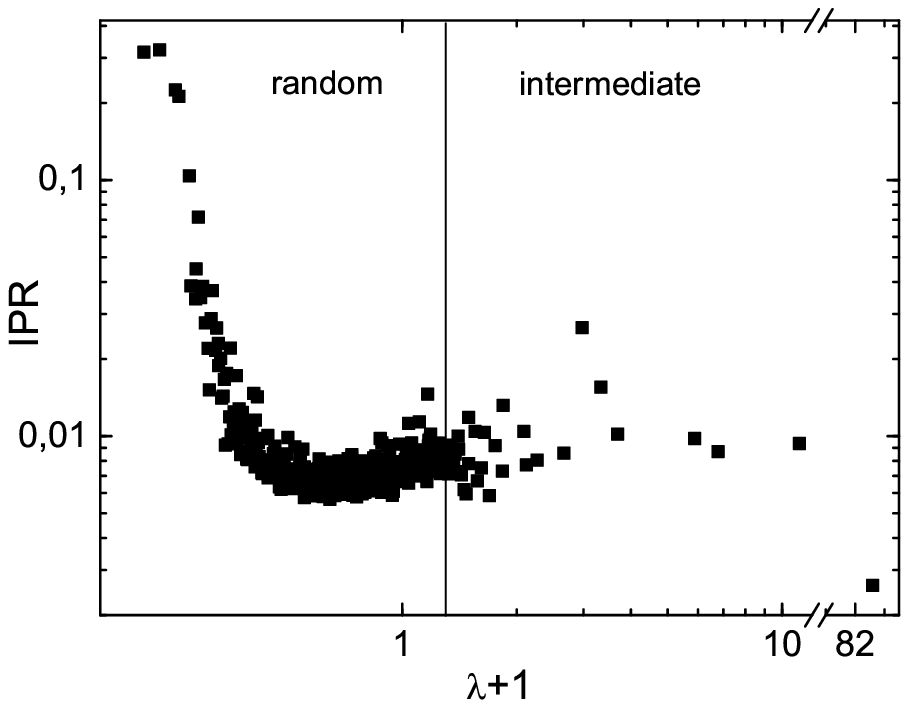}
\end{center}
\caption{Left: The spectral density of the weight matrix. The inset shows
  the random bulk and the analytical curve. Right: The IPRs of the
  eigenvectors as a function of the corresponding eigenvalue.}
\label{fig:dos}
\end{figure}

Fig. \ref{fig:dos} depicts the spectrum of the weight matrix together with the
analytical results for Wishart matrices
(Eq. (\ref{eq:MP})).\footnote{Note that the spectrum is shifted to the
  left by 1, due to Eq. (\ref{eq:CW}). In \cite{Burda_GW} an improved fit is suggested based on the random matrix theory of power law distributed variables. However, the minor difference in the fitting is irrelevant from our points of view.} 
The analytical curve is fitted by visual inspection using $\sigma$, i.e.
the variance of the effectively random part of the correlation matrix, as an
adjustable parameter. Best fit is obtained with $\sigma \approx 0.86$,
which, substituted into Eq. (\ref{eq:MP2}), yields $\lambda_{max}
\approx 0.3$.  However, many eigenvectors corresponding to eigenvalues
above this bound are to a large extent random and on the other hand,
some below this bound contain information \cite{Utsugi}. Therefore,
$\lambda_{max}$ can only be considered as a suggestive indicator of
the crossover region between the random and intermediate parts of the spectrum.

Plerou \emph{et al.} \cite{RMT_Stanley, Stanley_long} have suggested
the use of inverse participation ratios (IPR), defined for vector $\vec{v}$ as
\begin{equation}
I(\vec{v}) = \sum_{i} v_{i}^{\phantom{i} 4},
\end{equation}
in the identification of the information carrying eigenvectors.  The
idea behind this is that the more localized the eigenvector is, the higher is 
its IPR. From the right panel of Fig. \ref{fig:dos}, which depicts $I(\vec{v})$ 
for the eigenvectors of the weight matrix as a function of the
corresponding eigenvalue, we see that most of the random and
intermediate eigenvectors have similar IPRs (see also \cite{Utsugi}). Thus IPR
does not seem to be an efficient tool to distinguish the information carrying eigenvectors from the
rest.\footnote{The high IPRs of the lowest ranking eigenvectors are due to the
well known fact that they are localized to pairs of stocks with the
very highest correlation coefficients \cite{RMT_Stanley,Stanley_long,
  Utsugi}.} More sophisticated analysis is needed.

\subsection{Localization of the eigenvectors}  

\begin{figure}[!h]
\begin{center}
\includegraphics*[width=180pt]{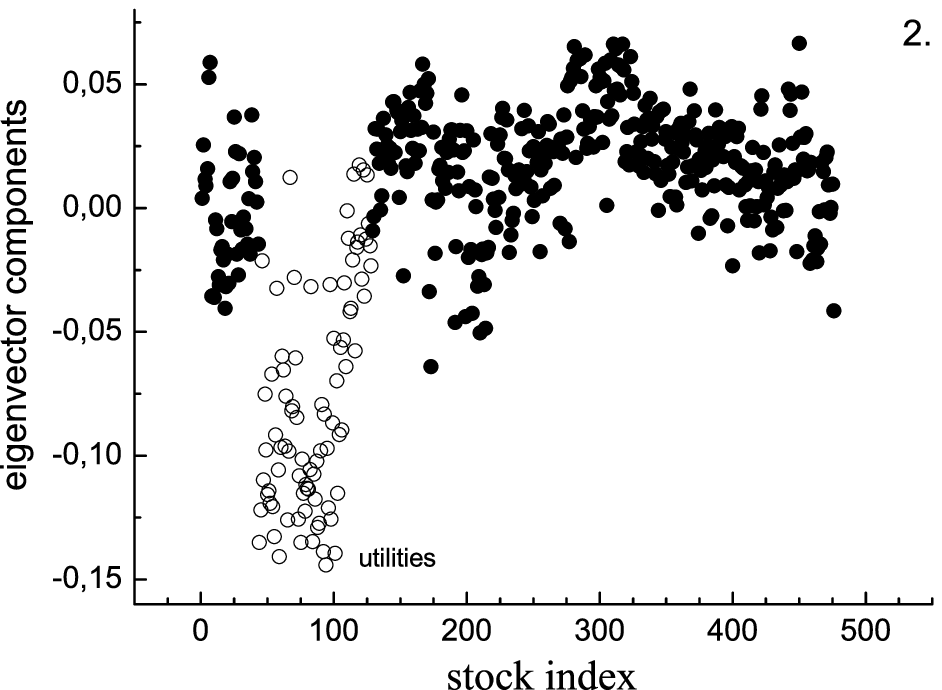}
\includegraphics*[width=180pt]{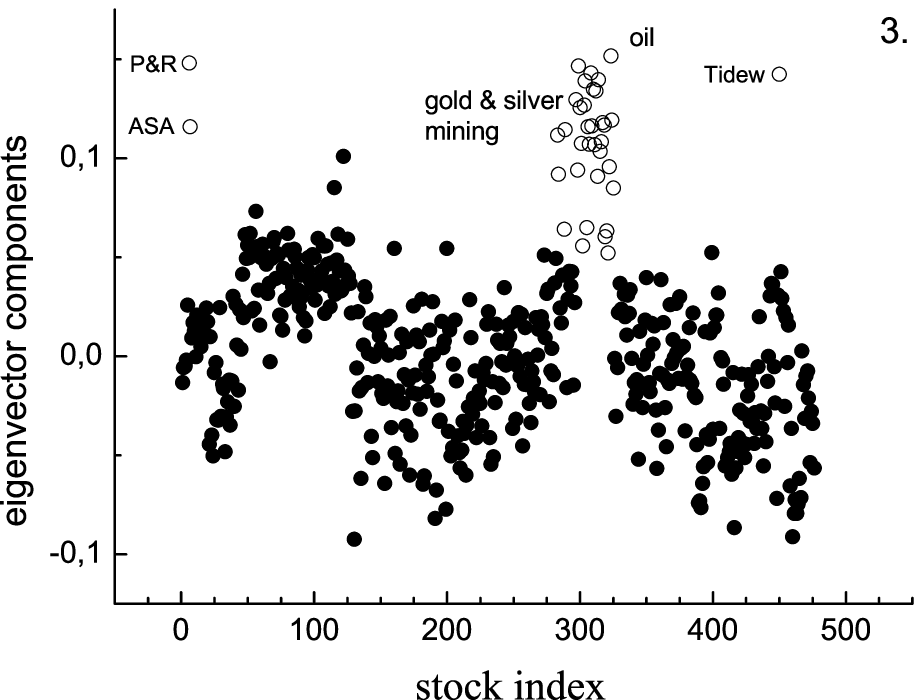}
\includegraphics*[width=180pt]{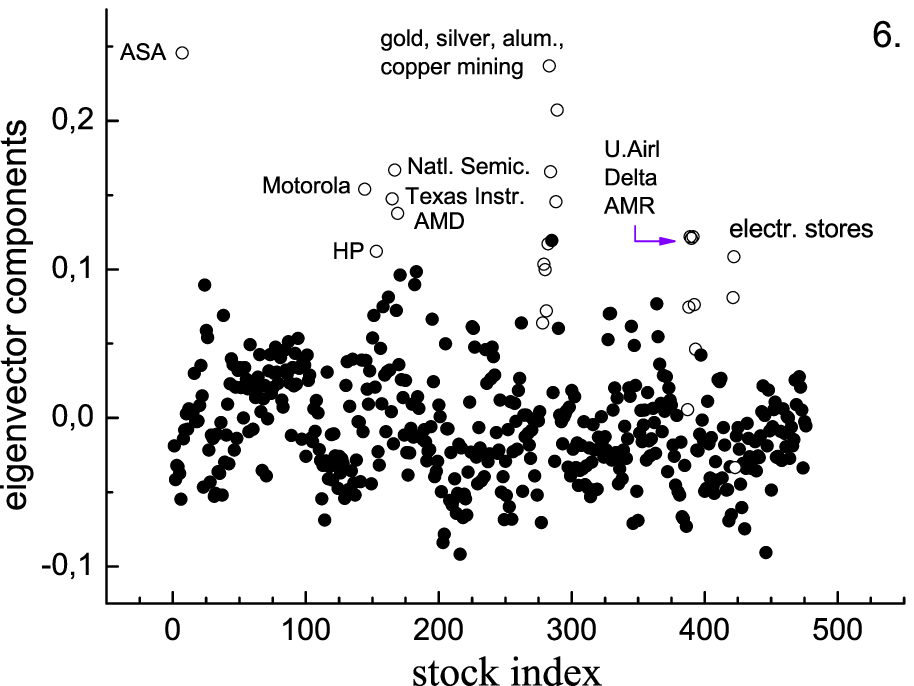}
\includegraphics*[width=180pt]{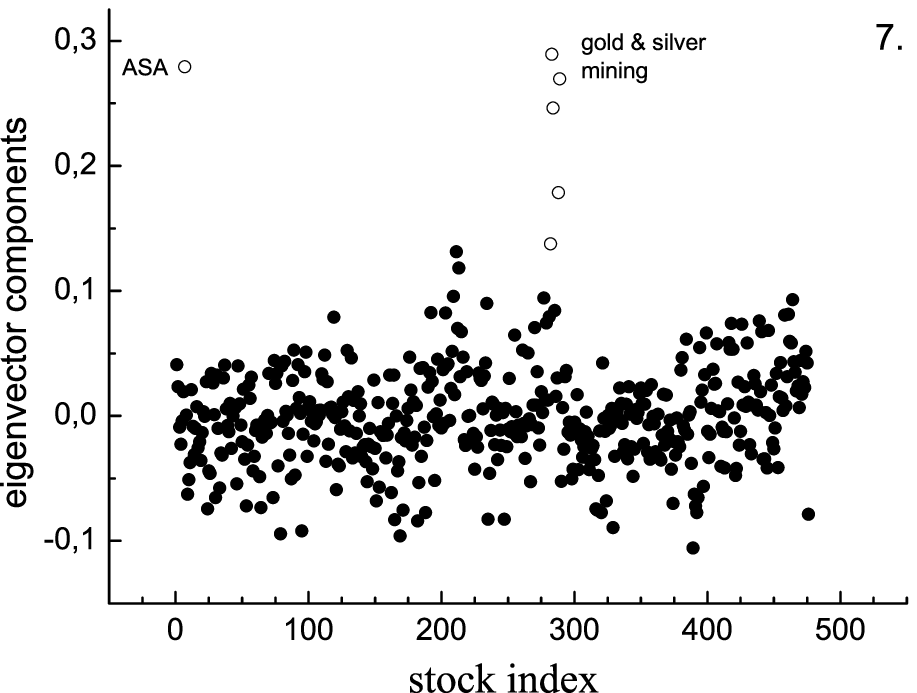}
\includegraphics*[width=180pt]{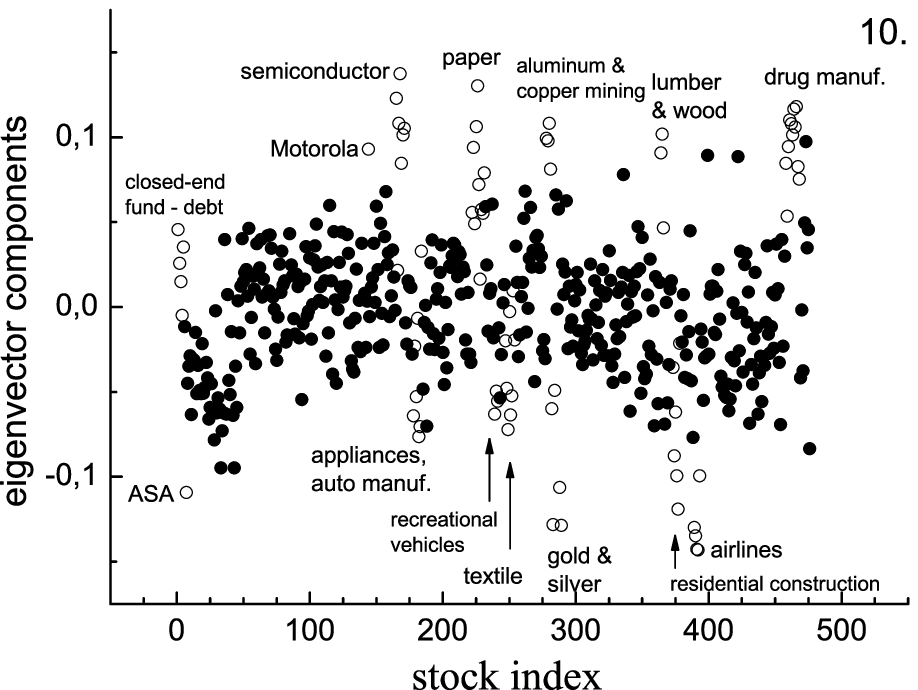}
\includegraphics*[width=180pt]{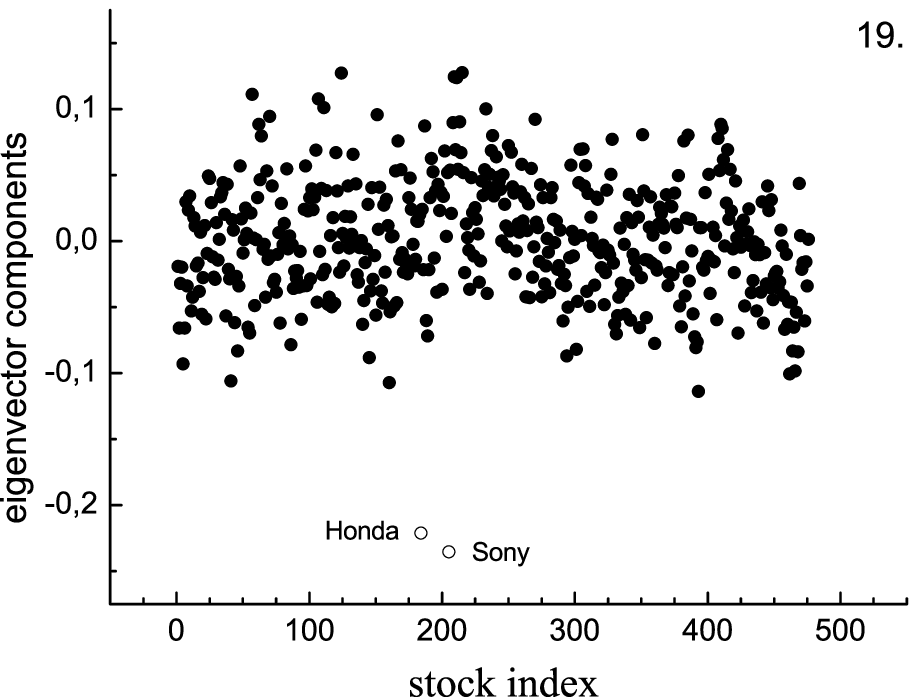}
\end{center}
\caption{Component sizes of chosen eigenvectors.  The number in each
  panel indicates the rank of the corresponding eigenvalue. Horizontal
  ordering is such that stocks belonging to same business sectors are
  next to each other and open symbols are used as a guide to the
  eye. For abbreviations, see text.}
\label{fig:components}
\end{figure}

We have seen that the gradual increase of the noise content aleady
makes the identification of the clusters in the network a difficult
task. Here we will go into the further difficulties caused be the
complexity of the localization of the information carrying
eigenvectors. Financial correlations are particularly appropriate to
investigate this point as independent classification schemes exist to
compare with. In the following we will take advantage of this
information in the example-like analysis of a couple of interesting
eigenvectors.  The components of the eigenvectors studied are illustrated 
in Fig. \ref{fig:components}, in which the (horizontal) ordering of the 
stocks is such that stocks belonging to same business sector according 
to Yahoo classification \cite{yahoo} are next to each other. This makes 
the eigenvectors localized on a business sector stand out more clearly.   

The highest ranking intermediate eigenvector, namely the second
eigenvector is a good example of an eigenvector localized on a business
sector. The components corresponding to the utilities sector stand out very
cleanly in Fig. \ref{fig:components}. One should notice, however, that
this would not be the case without the chosen horizontal ordering of
the companies. Without a priori information we would not be able to define 
boundaries for this cluster.   

The third eigenvector, which is mainly localized
on oil and gold \& silver mining companies is already a more difficult one. 
The other large components of this eigenvector correspond to 
Petroleum \& Resources (P\&R), a financial company specialized in the energy 
sector, Tidewater (Tidew), which provides vessels and services for the offshore
energy industry, and ASA Ltd. (ASA), an investment company interested
in precious metal mining. The thresholding analysis (not presented here) 
shows that the companies corresponding to the largest components of
this eigenvector form two clusters. The third eigenvector is not the only
example of a high ranking intermediate eigenvector localized on more
than one cluster. The sixth eigenvector, for example, is localized on
gold \& silver mining, leading electronics manufacturers \& electronics
stores, and air transportation companies.  Again, the thresholding analysis
shows that all these industries form their own clusters. Interestingly 
the seventh eigenvector is localized solely on the gold \& silver
mining-related companies. 
%Further analysis of the eigenvectors reveals that each of these clusters have
%other eigenvectors (the seventh, the 11th and the 12th respectively)
%more or less localized on it. 

One encounters further difficulties, when analyzing e.g. the tenth eigenvector, 
which has a very complex structure. As illustrated in Fig. \ref{fig:components}, 
it is localized on a large number of industry branches, most of which can be 
found by the thresholding analysis. However, without some prior information, 
interpretation of this eigenvector is impossible. On the other hand, surprisingly, 
the 19th eigenvector can be straightforwardly interpreted although the 
corresponding eigenvalue is close to the random part of the spectrum 
($\lambda_{19} \approx 0.55$) and 
the neighbouring eigenvectors are to a large extent random. This eigenvector
is strongly localized on Sony and Honda, the only Japanese companies in the data 
set. It should be noted that several eigenvectors corresponding to 
the lowest ranking eigenvalues are localized on pairs of companies
with highest correlation coefficients. 

To summarize, it seems evident that the cluster structure of a network
cannot be easily deduced from the eigenvectors of the weight
matrix. Especially, interpretation of a single eigenvector 
is even more difficult than suggested in recent literature. Most of the 
information about the cluster structure can only be found by combining 
information from different eigenvectors\footnote{This was also suggested in
  \cite{Stanley_sectors}, in which symmetric and antisymmetric
combinations of eigenvectors are analyzed.}. There is, however, no
rule to tell, which linear combination of the eigenvectors should be
taken. Therefore, the extraction of the cluster structure from the
eigenvectors without a priori knowledge about the nodes (here
companies) seems to be a too formidable task.

\subsection{Diffusion based approach}

\begin{figure}[!h]
\begin{center}
\includegraphics*[width=250pt]{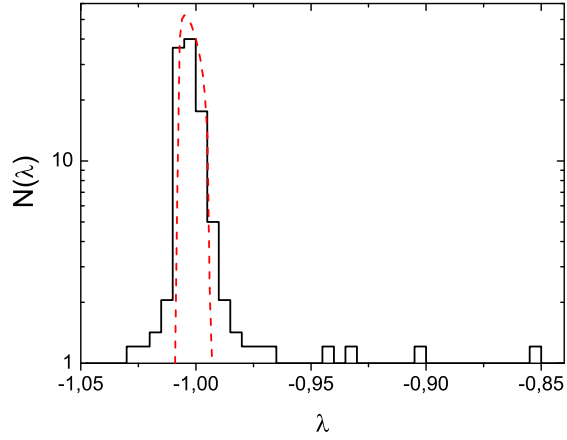}
\end{center}
\caption{(Color online) Spectral density of the diffusion matrix
  corresponding to the larger dataset (solid line), and the average
  spectral density over a system of $1000$ random references (dashed line). The
  trivial eigenvalue at $0$ is not shown.}
\label{fig:dosdiff}
\end{figure}

\begin{figure}[!h]
\begin{center}
\includegraphics*[width=250pt]{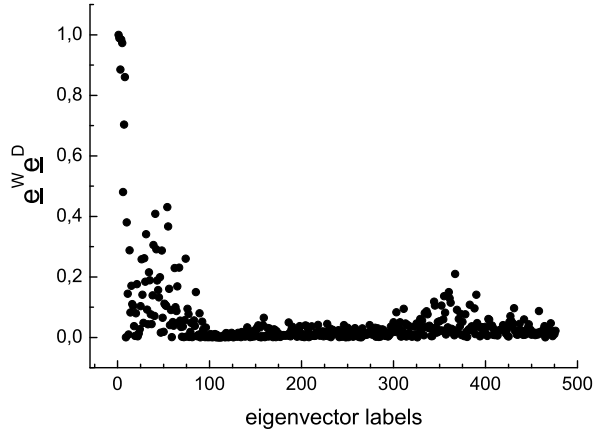}
\end{center}
\caption{Scalar products of the 
  eigenvectors of the diffusion matrix and the corresponding eigenvectors
 (i.e. eigenvectors with similar localization) of the weight
  matrix. The eigenvectors of the diffusion matrix are ordered
  according to decreasing rank (x-axis).}  
\label{fig:simil}
\end{figure}

In section \ref{sec:eigvec} we reasoned, how the cluster structure of
a network should affect the diffusion process. The spectrum of the
diffusion matrix (depicted by the solid line in
Fig. \ref{fig:dosdiff}) has, as expected, similar structure with that
of the weight matrix. For diffusion
matrices, results corresponding to Eqs. (\ref{eq:MP}) and (\ref{eq:MP2}) are 
not known, but a random reference function can be obtained numerically by 
constructing diffusion matrices from random weight matrices generated 
with the method presented in section \ref{sec:first_emp} and in \cite{Pafka}
(dashed line in Fig. \ref{fig:dosdiff}).
In Fig. \ref{fig:simil} we compare the eigenvectors of the weight and 
diffusion matrices. We see that the highest ranking eigenvectors of the 
diffusion matrix are very close to the corresponding eigenvectors 
(i.e. eigenvectors with similar localization) of the weight
matrix. Their distance increases when the random part of the spectrum
is approached and the correspondence between pairs of eigenvectors looses its meaning. 

\begin{figure}[!h]
\begin{center}
\includegraphics*[width=250pt]{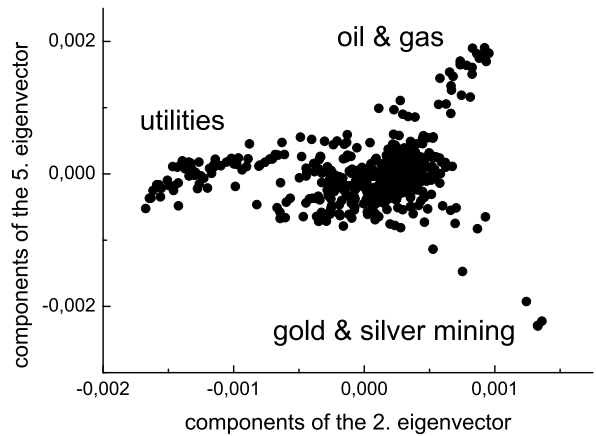}
\end{center}
\caption{Components of the fifth eigenvector of the normal
  matrix as a function of the components of the second eigenvector.}
\label{fig:starplot}
\end{figure}

The analysis of the eigenvectors of the normal matrix is again
non-trivial. Naturally, there are correlations between the components of
different eigenvectors, but it is impossible to identify
clusters without a priori information. Best results in two dimensions 
were obtained with the eigenvectors of ranks two and five 
(see Fig. \ref{fig:starplot}). Visual inspection of this plot allows us 
to identify the oil \& gas, utilities and gold \& silver mining clusters, 
although the determination of the boundaries is again difficult.

\section{Asset graph approach to the clustering of stocks}
\label{sec:assetg}

\begin{figure}
\begin{center}
\includegraphics[width=160pt]{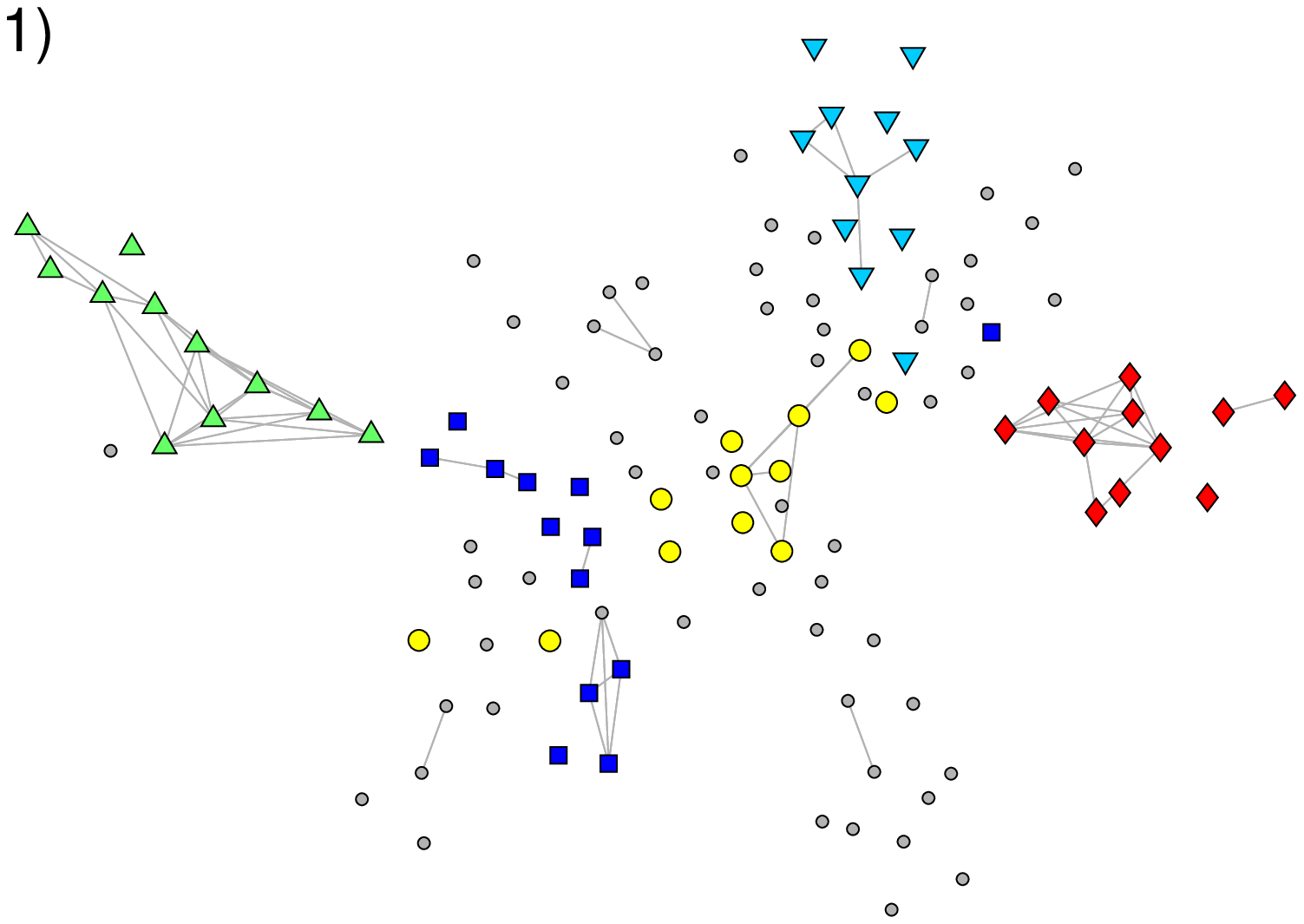}
\includegraphics[width=160pt]{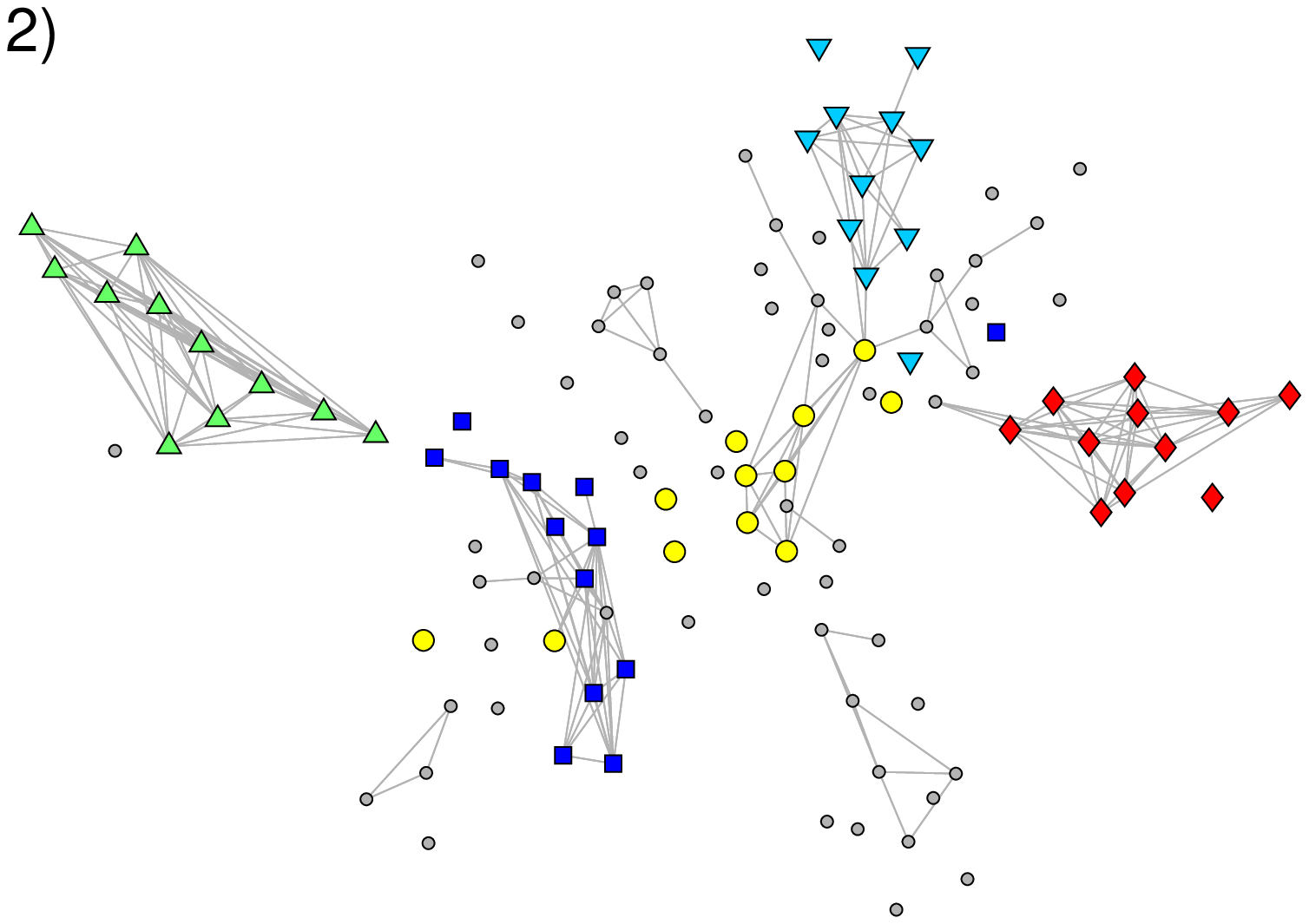}
\includegraphics[width=160pt]{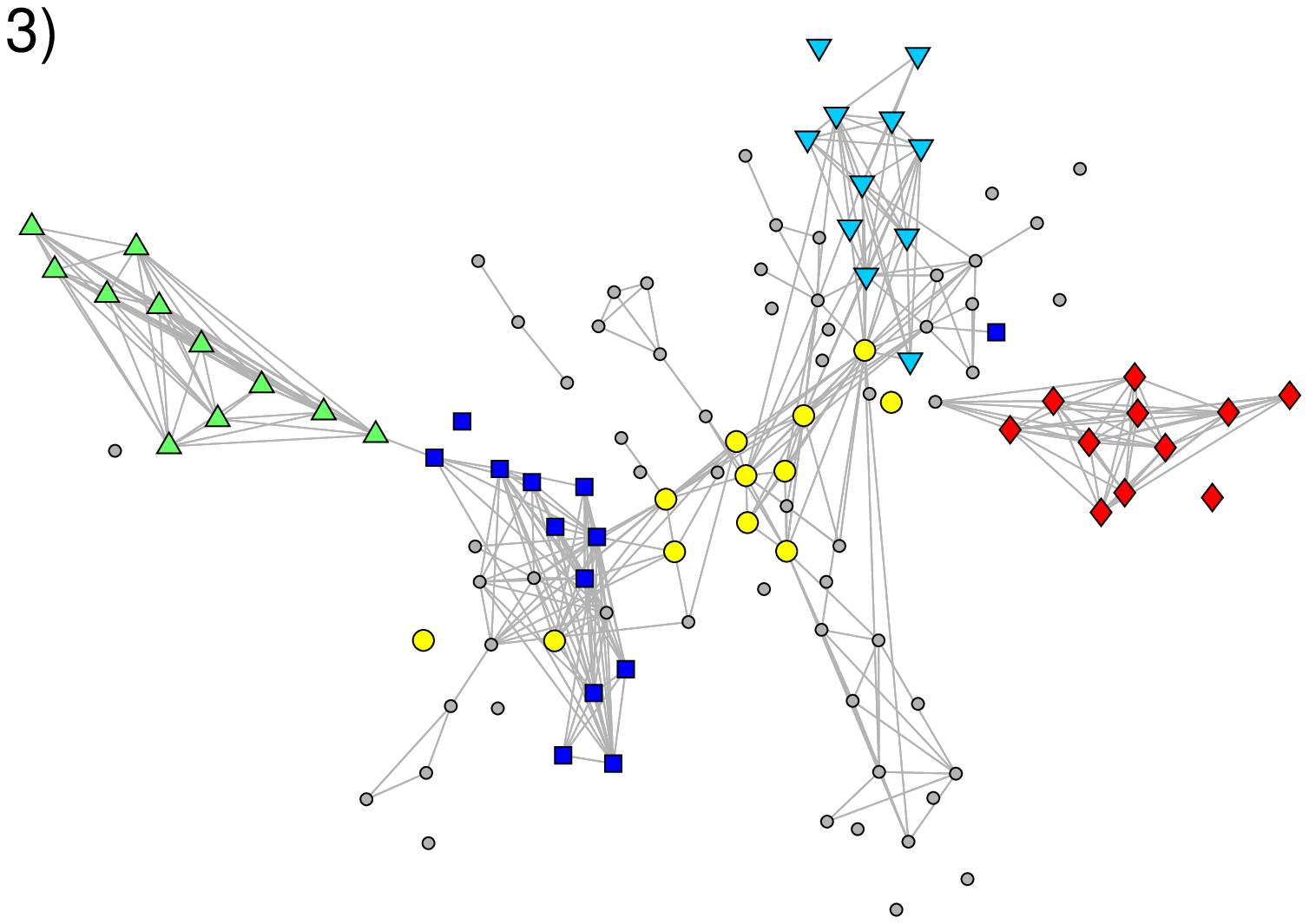}
\includegraphics[width=160pt]{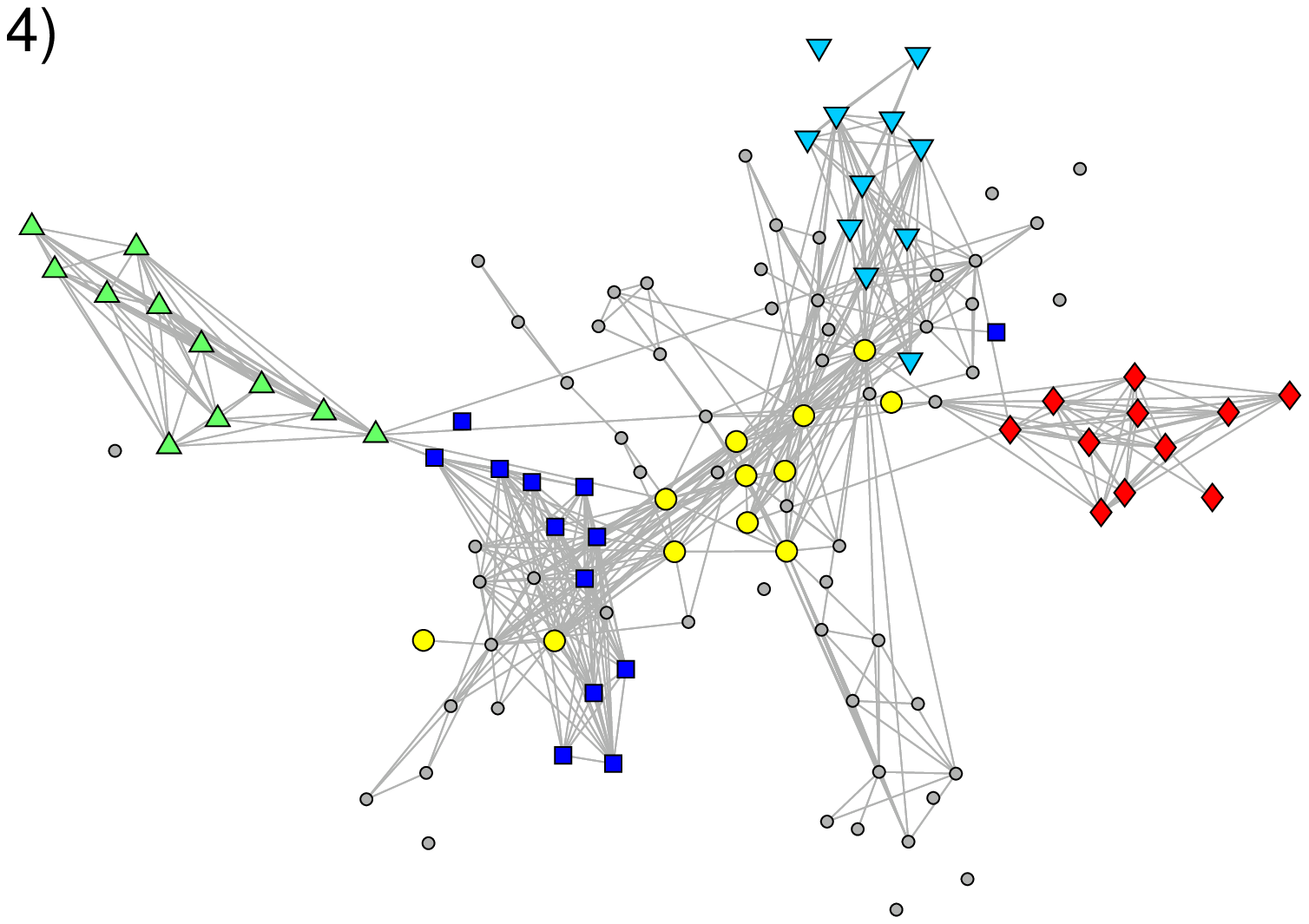}
\caption{(color online) The asset graph constructed using the full
  correlation matrix $\boldsymbol{C}$
  for link occupation values 1) $p=0.01$, 2) $p=0.03$, 3)
  $p=0.05$ and 4) $p=0.07$. Forbes classification~\cite{forbes} has been
  used and companies belonging to Energy
  sector are denoted by \textcolor{green}{$\blacktriangle$}, Electric Utilities
  industry by \textcolor{red}{$\blacklozenge$}, Healthcare sector by
  \textcolor{cyan}{$\blacktriangledown$}, 
  Basic Materials sector by \textcolor{blue}{$\blacksquare$} and
  Finacial as well as Conglomerates
  sector by \textcolor{yellow}{$\bullet$}. Other nodes are denoted by $\bullet$.}
\label{fig:assetgs}
\end{center}      
\end{figure}

In this section we study the clustering of stocks using asset
graphs~\cite{onnela:clust}.\footnote{In this section, only the smaller
  dataset is studied}   
An asset graph is constructed by ranking the non-diagonal
elements of the correlation matrix and adding links between stocks
one after the other, starting from the strongest correlation
coefficient. The network thus emerging can be characterized by a parameter
$p$, which is the ratio of the number of added links to the number of all
possible links, $N(N-1)/2$. Asset graphs constructed using the full
correlation matrix $\boldsymbol{C}$ are illustrated\footnote{For illustration, the node
coordinates are generated with Pajek by plotting the MST.} in Fig. \ref{fig:assetgs}
for link occupation values of $p=0.01$, $p=0.03$, $p=0.05$, and $p=0.07$.
An immediate observation is that some clusters stand out very
cleanly and can already be identified by visual inspection. These
clusters correspond very well to business sectors and industries
according to Forbes classification~\cite{forbes}. However,
we cannot expect to find all clusters this way and for large $N$ this
approach cannot be applied. It is clear that we need more
sophisticated methods. One possibility, suggested by Onnela \emph{et al.}, 
is to define a cluster as an isolated component and study the evolution 
of these components as a function of $p$. Another possibility is to 
apply some known community detection method for binary graphs (see \emph{e.g.}
\cite{newman2,CliquePerc,Reichardt,reichardt2,newman1}) as a function
of $p$. However, the problem with these approaches is that there is no global 
threshold value of $p$ with which we would find (almost) all the information 
about the clusters. A more comprehensive picture about the cluster structure can be 
obtained by studying the evolution of each cluster separately as a function of
$p$. Alternatively, we can also define our asset graphs in a different way.

\begin{figure}
\begin{center}
\includegraphics[width=160pt]{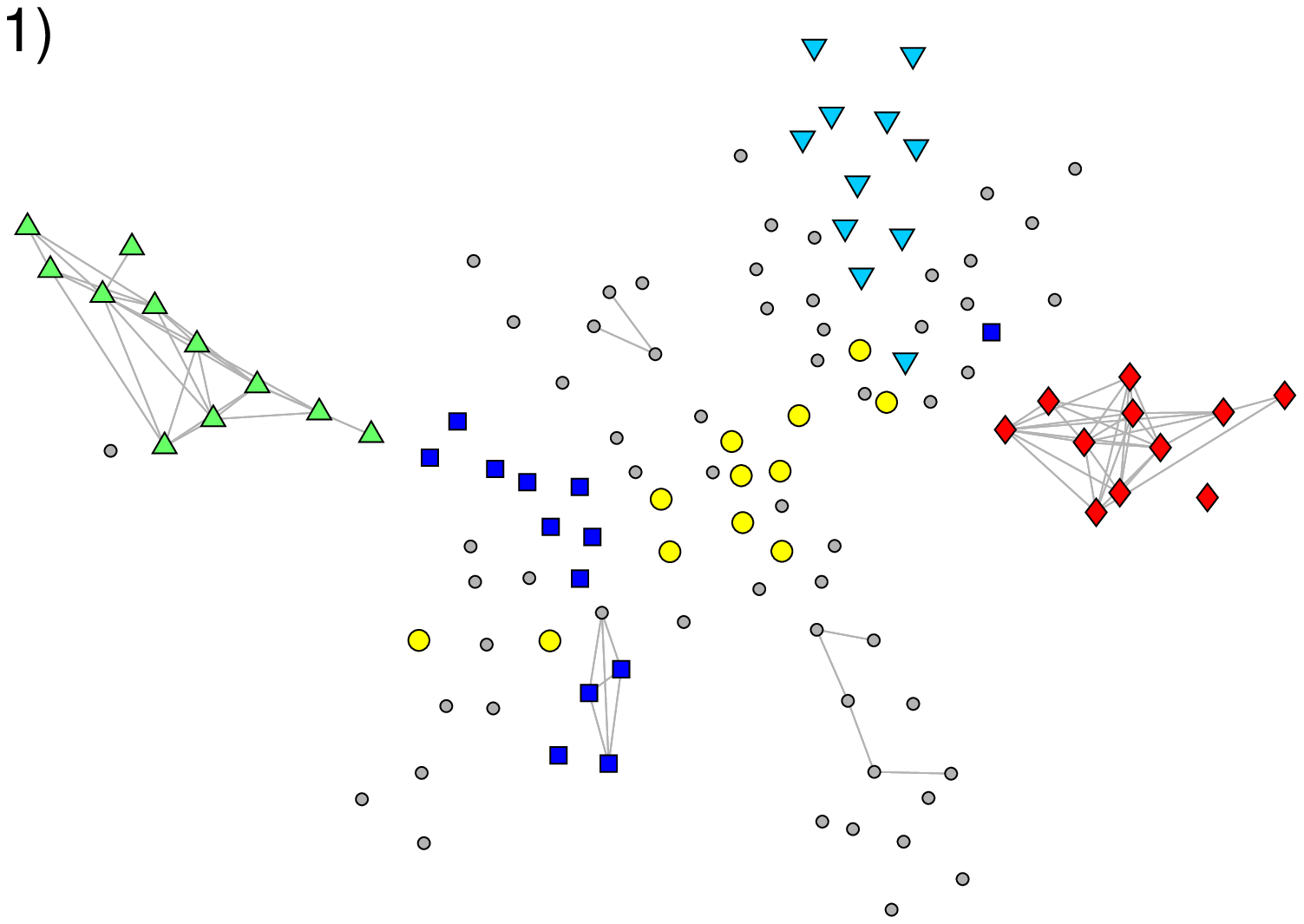}
\includegraphics[width=160pt]{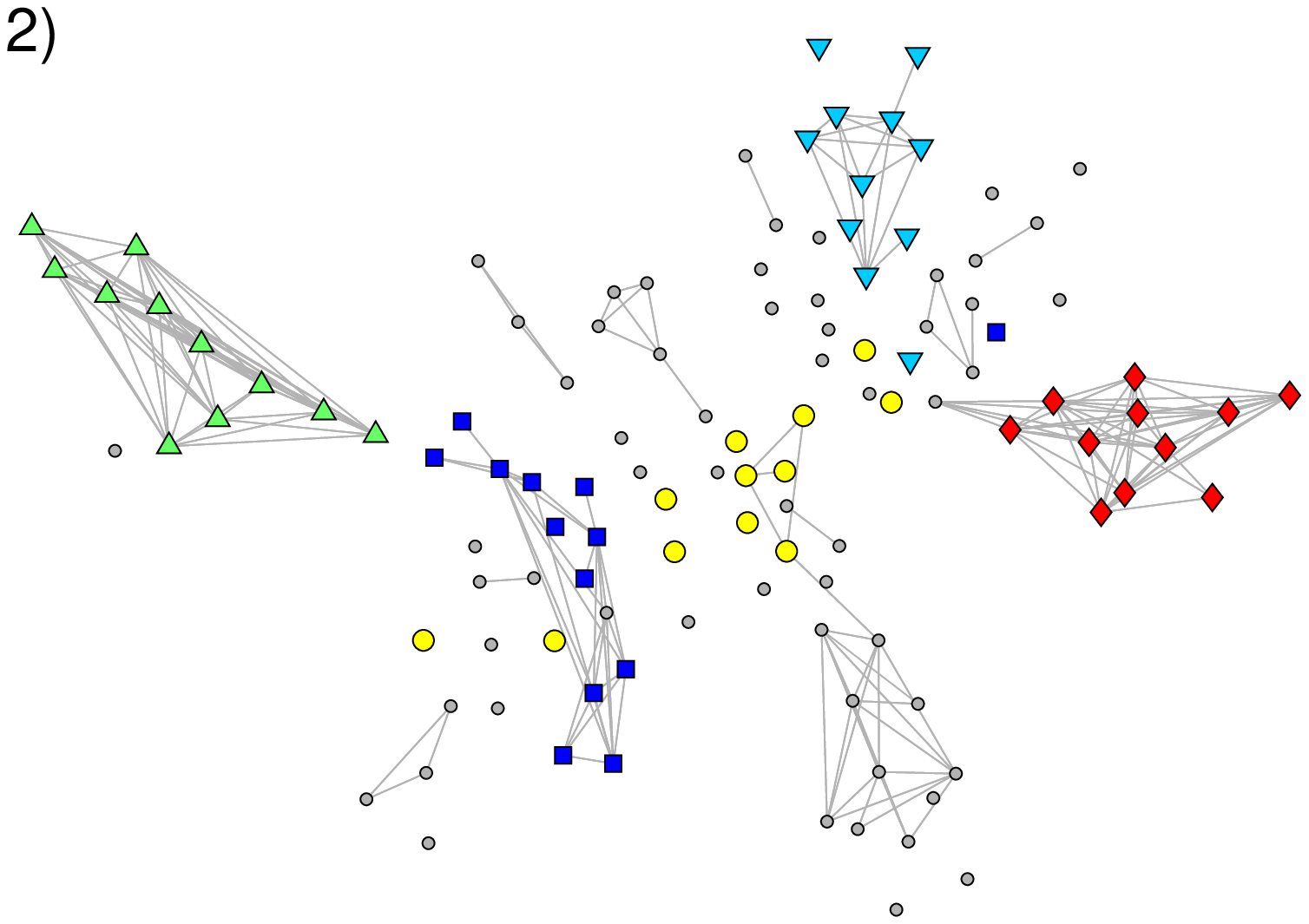}
\includegraphics[width=160pt]{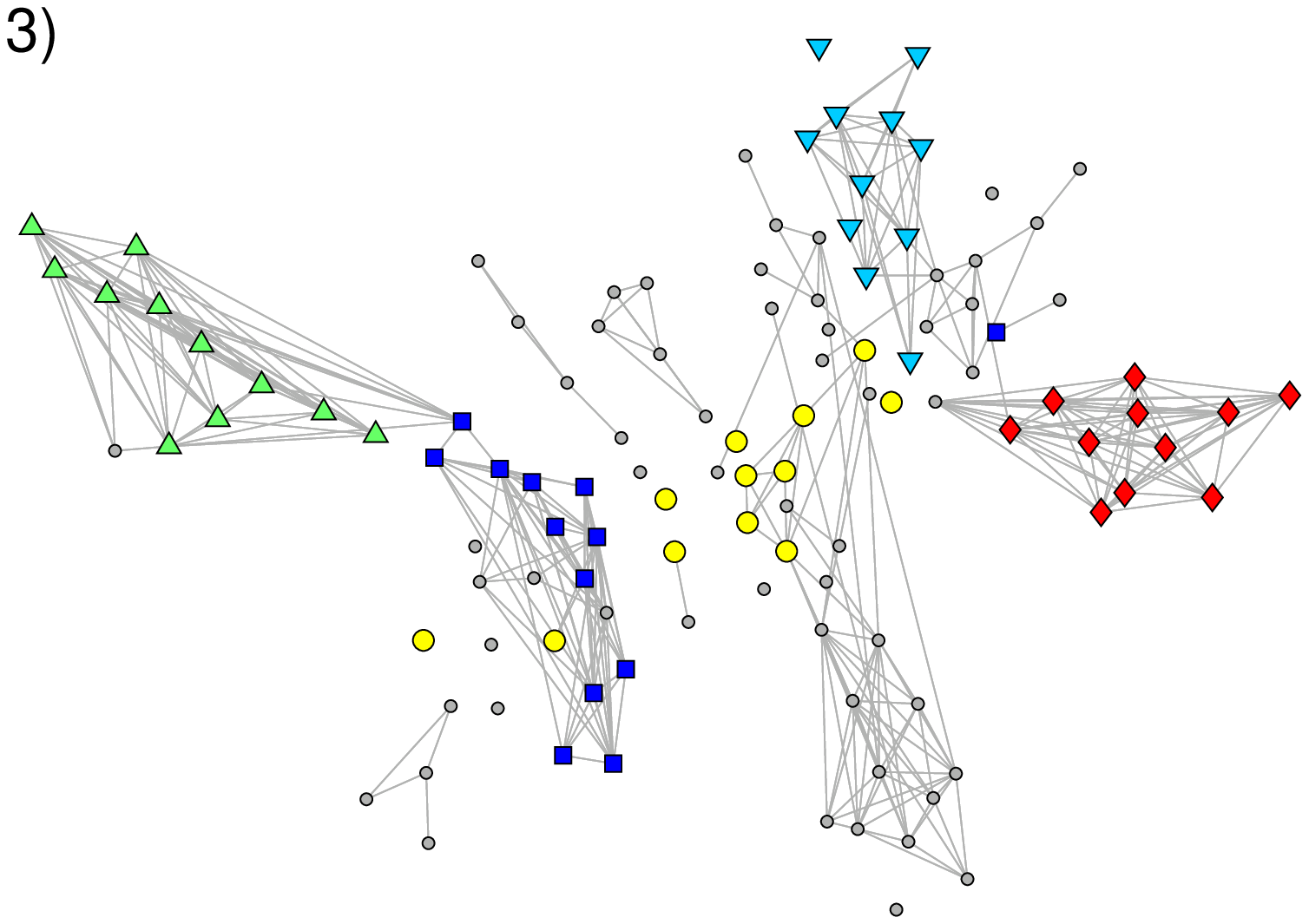}
\includegraphics[width=160pt]{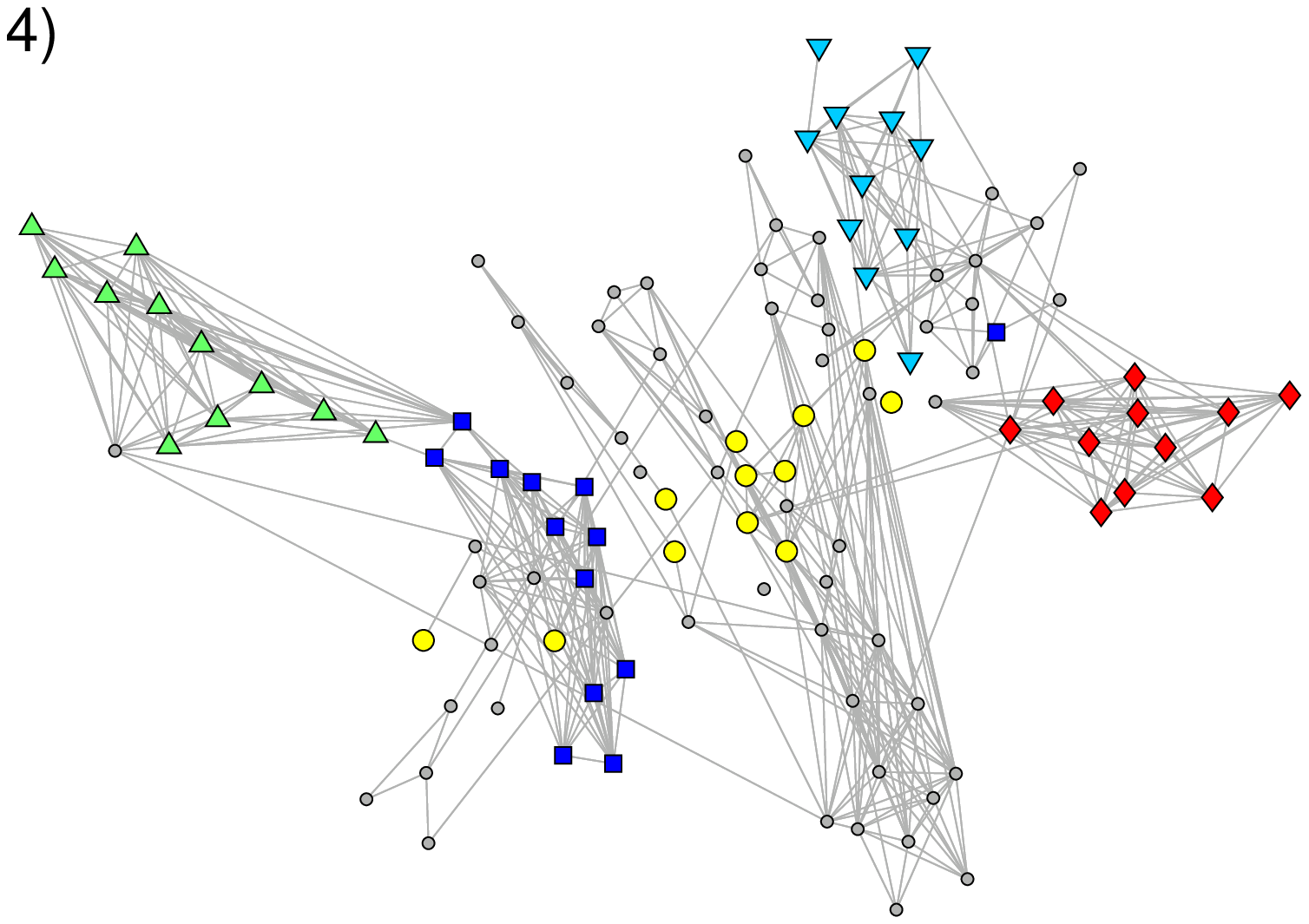}
\caption{(color online) The asset graph constructed using
  $\boldsymbol{C}_{-m}$ (i.e. correlation matrix from which the effect of
  the market eigenpair has been filtered out) for link 
  occupation values 1) $p=0.01$, 2) $p=0.03$, 3) $p=0.05$ and 4) $p=0.07$. Nodes are
  denoted as in Fig.~\ref{fig:assetgs}.}
\label{fig:assetg_filts115}
\end{center}      
\end{figure}

\begin{figure}
\begin{center}
\includegraphics[width=180pt]{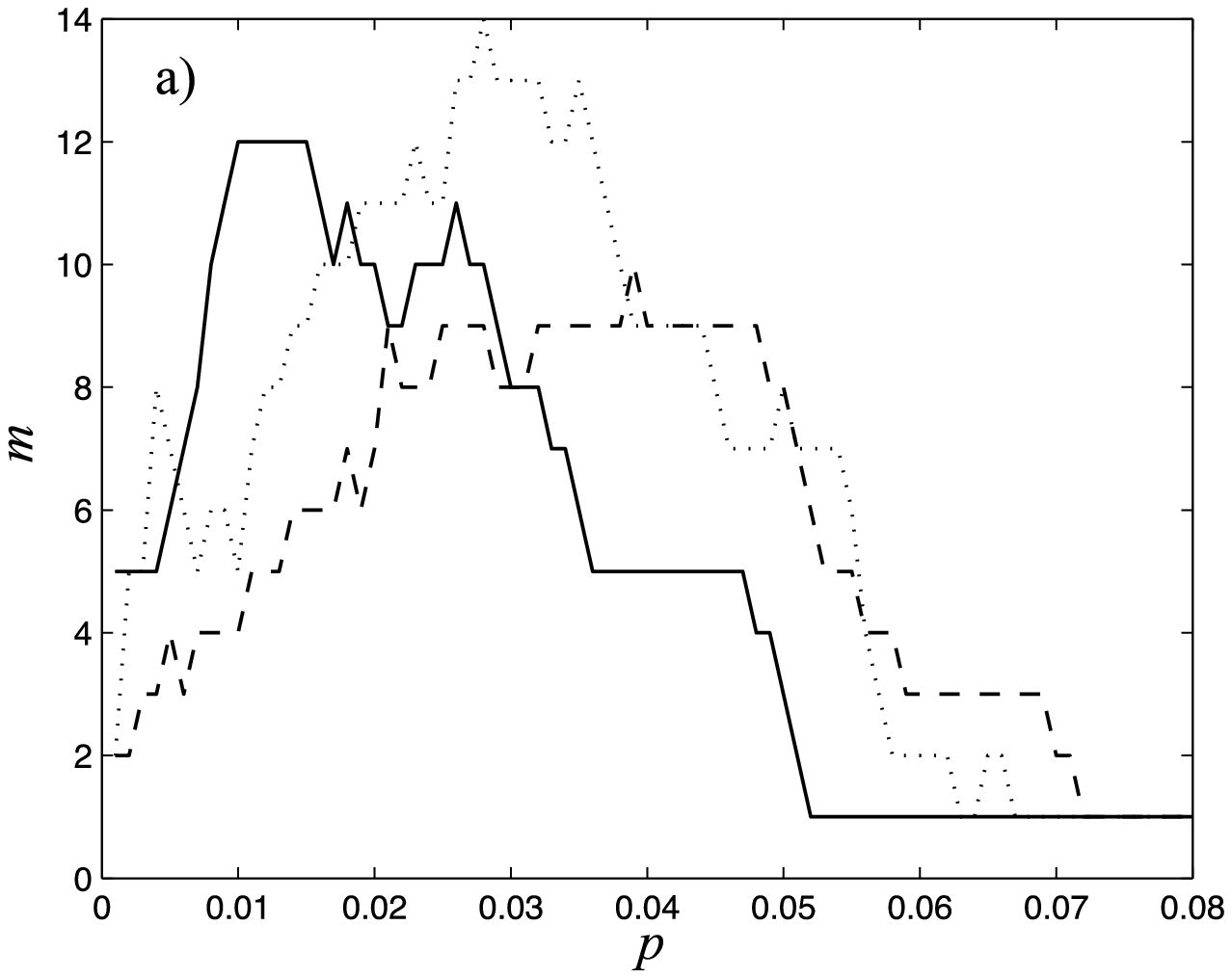}
\includegraphics[width=180pt]{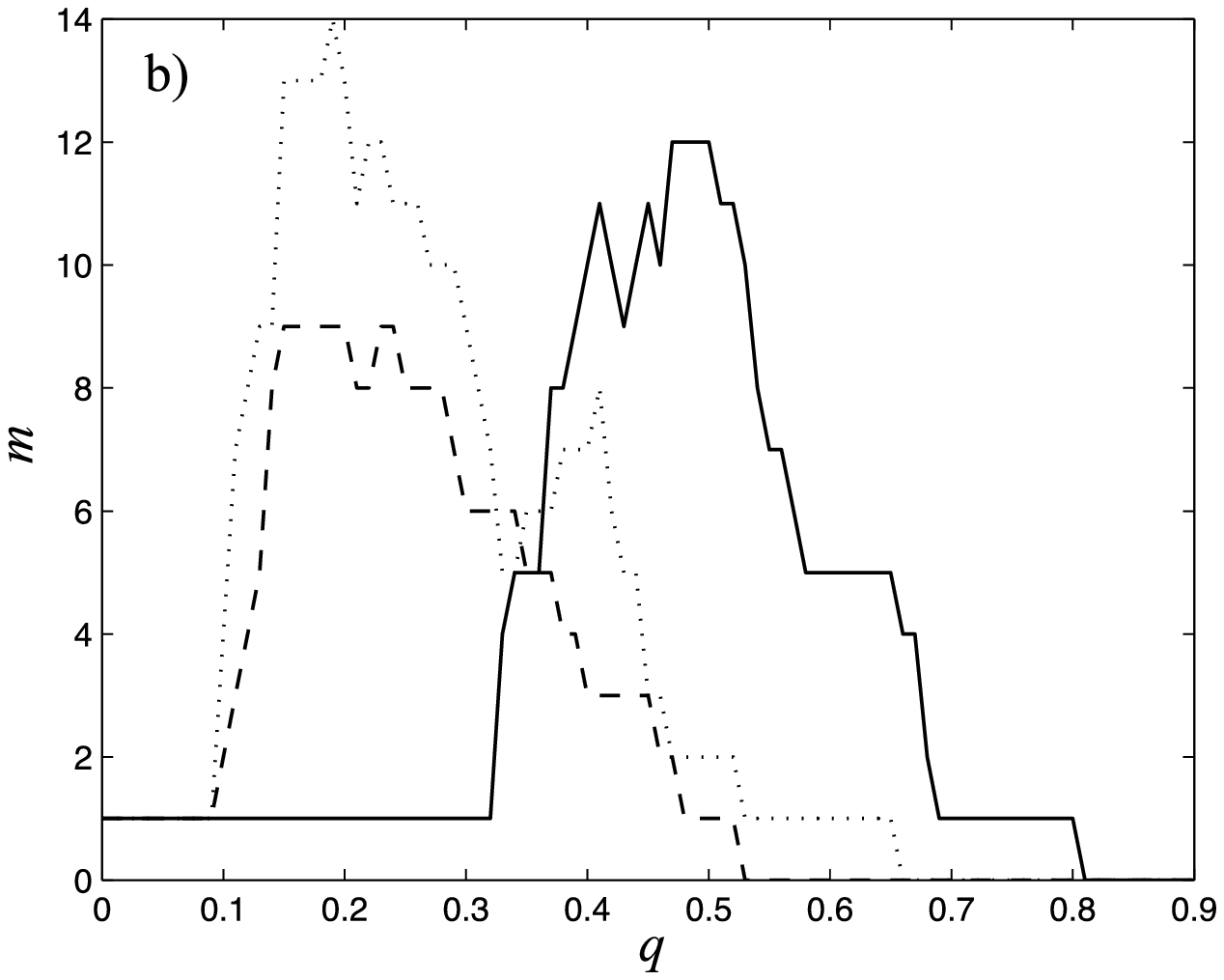}
\caption{The number of isolated components $m$ as a function of $p$
  (panel a) and $q$ (panel b) in
  asset graphs constructed from $\boldsymbol{C}$ (solid line), $\boldsymbol{C}_{g}$ (dashed
  line) and $\boldsymbol{C}_{-m}$ (dotted line). For
  $\boldsymbol{C}_{g}$ and $\boldsymbol{C}_{-m}$ there is a sudden
  increase at $q_{c}\approx 0.1$ (panel b). This jump, however,
  is not seen in panel a. (Notice that the number of links increases as
  a function of $p$ and decreases as a function of $q$.)}
\label{fig:number_p}
\end{center}      
\end{figure}

The largest components of the market eigenvector, most\-ly conglomerates
and financial companies, have significant correlations with almost all the
other companies. This leads to the phenomenon clearly seen in
Fig. \ref{fig:assetgs} that different clusters in asset graphs merge
mostly through nodes corresponding to these companies (for further
discussion see~\cite{Jeong}). Therefore, it is interesting to study asset
graphs without the effect of the market eigenvector. This can be done
by expanding the correlation matrix as 
\begin{equation}
\boldsymbol{C}=\sum_{i=1}^{N} \lambda_{i} |e_{i} \rangle \langle e_{i}|,
\end{equation}
where the eigenvalues are sorted according to decreasing rank and
constructing the asset graphs using the matrix defined by
\begin{equation}
\boldsymbol{C}_{-m}=\sum_{i=2}^{N} \lambda_{i} |e_{i} \rangle \langle e_{i}|.
\end{equation}
These are illustrated in Fig.~\ref{fig:assetg_filts115} for
link occupation values of $p=0.01$, $p=0.03$, $p=0.05$, and $p=0.07$. 
Here the most significant difference compared to Fig. \ref{fig:assetgs} is that, 
since the market eigenvector is excluded, the degrees of the nodes with the 
highest betweeness centralities in the MST (see Fig. $1$ in \cite{Tapio_Tokyo}) 
are much lower and some components remain isolated for larger values of $p$. 
However, from panel a of Fig.~\ref{fig:number_p}, where we illustrate as a 
function of $p$ the number of isolated components of size larger than
one, as well as from Fig.~\ref{fig:assetg_filts115} we notice that the problem 
still stands. There is no global threshold value of $p$ that would reveal 
all the clusters.

Kim \emph{et al.}~\cite{Jeong} have approached the problem by
defining, what they call the group correlation matrix by
\begin{equation} \label{eq:kim}
\boldsymbol{C}_{g} = \sum_{i=2}^{N_{g}} \lambda_{i} |e_{i} \rangle \langle e_{i}|,
\end{equation}
where $N_{g}$ is used to exlude the effect of the random eigenpairs.
From the previous section we know that by choosing $N_{g} < N$ we lose
some information, but the idea here is to get rid of most of the noise
without losing too much information. $N_{g}$ can be approximated by
comparing the eigenvalues to the theoretical eigenvalue density for 
random correlation matrices and by studying the localization of the 
eigenvectors. In the following we have used $N_{g}=10$.
\begin{figure}
\begin{center}
\includegraphics[width=160pt]{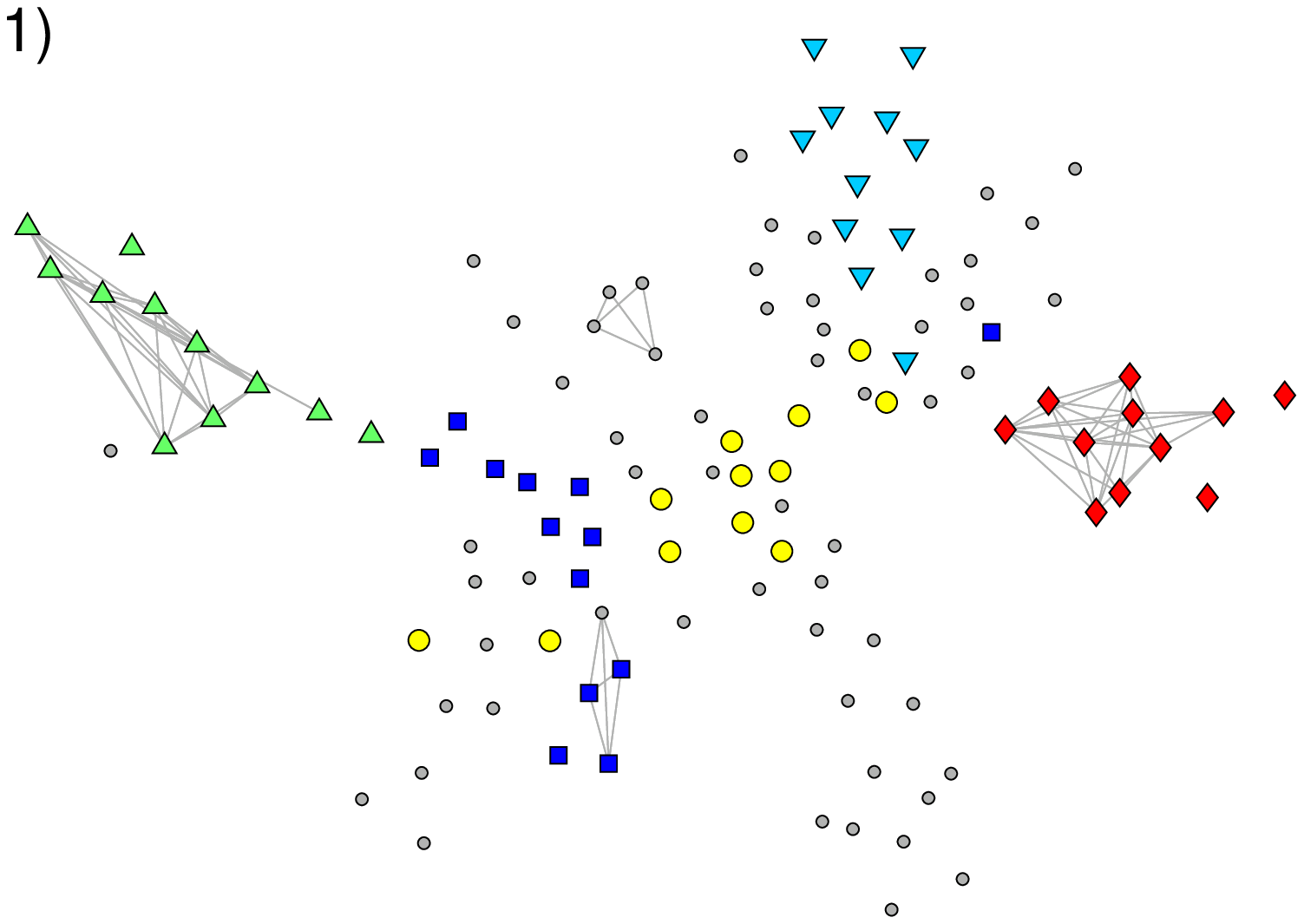}
\includegraphics[width=160pt]{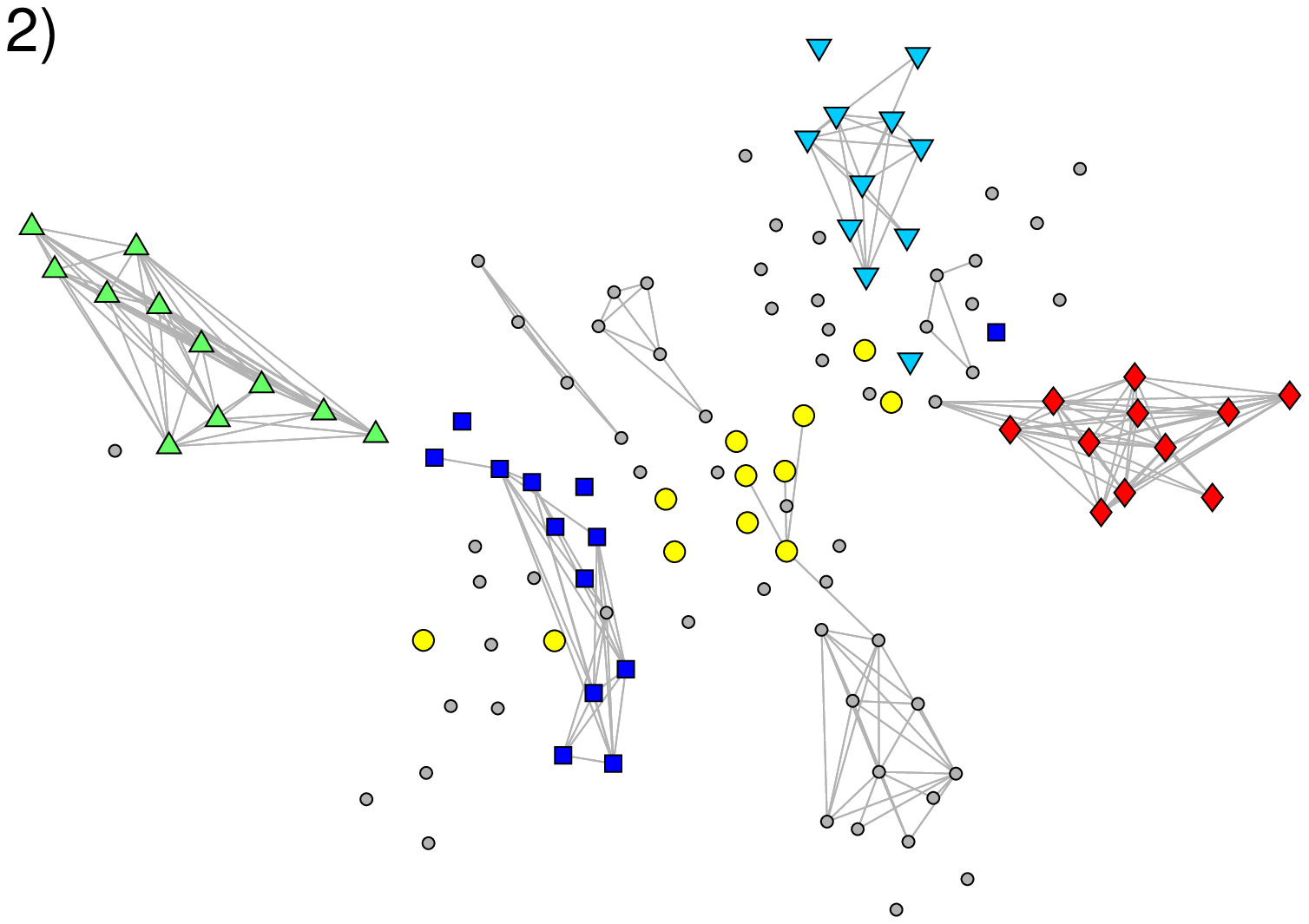}
\includegraphics[width=160pt]{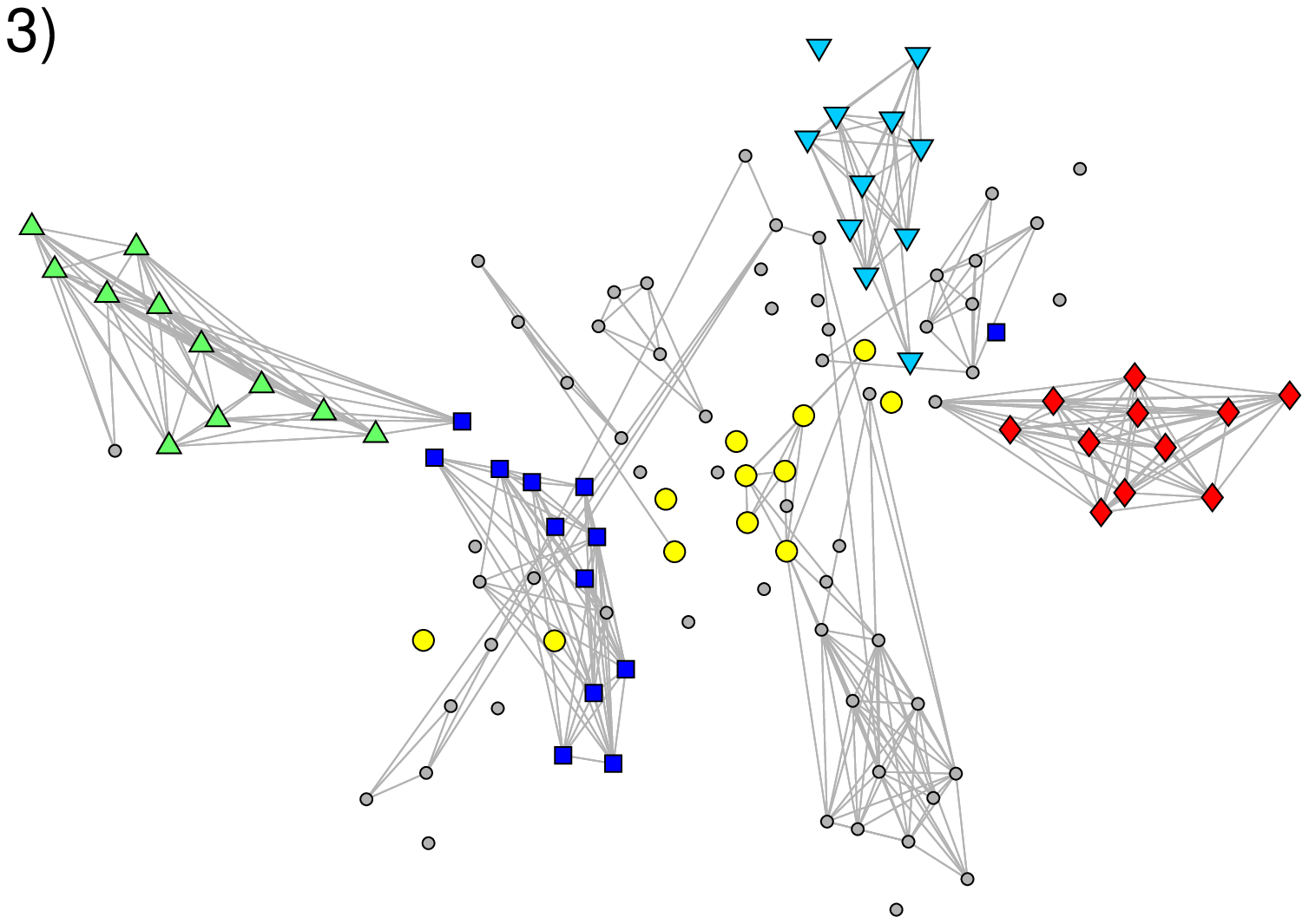}
\includegraphics[width=160pt]{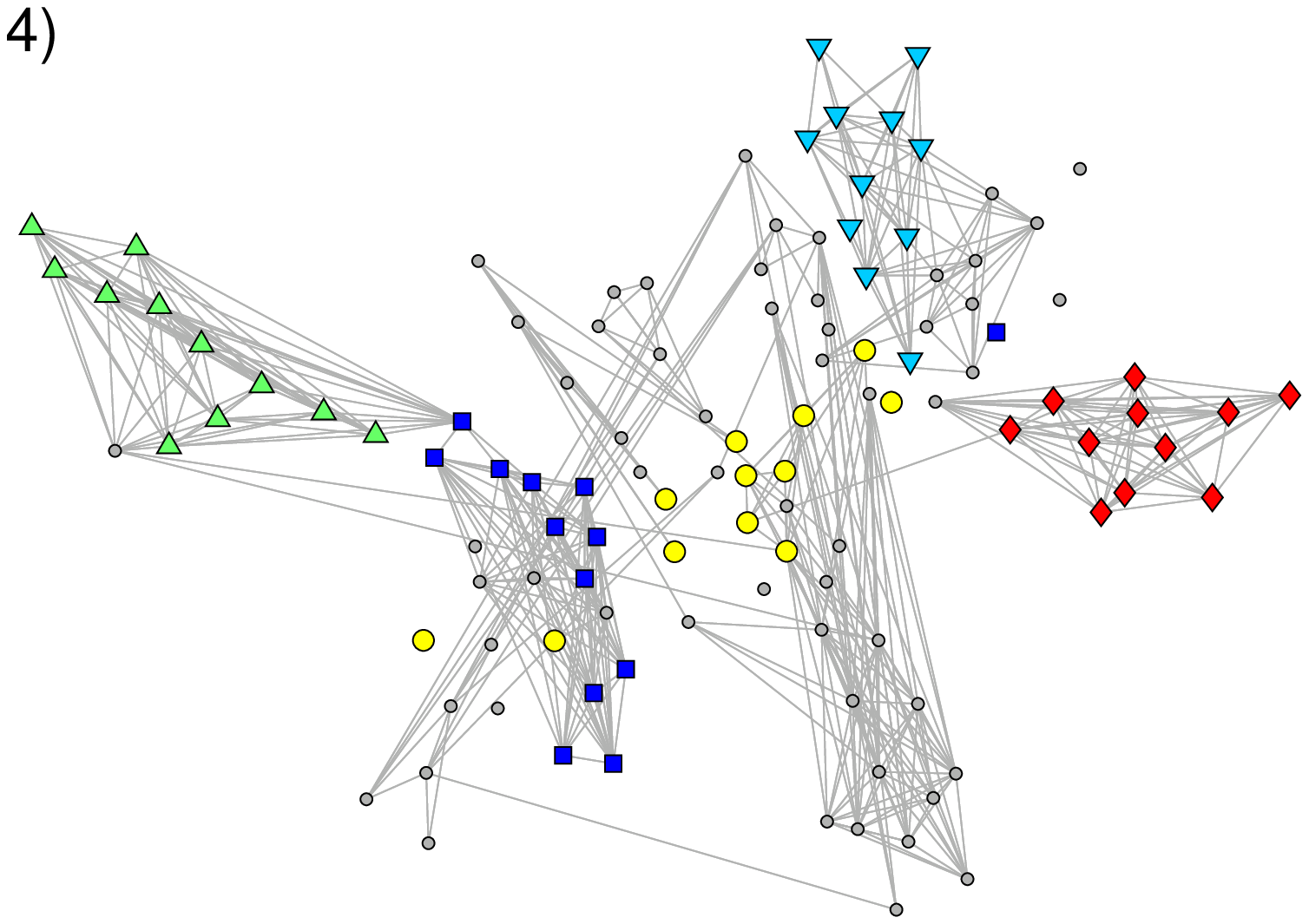}
\caption{(color online) The asset graph constructed using
  $\boldsymbol{C}_{g}$ (i.e. correlation matrix from which the effects
  of the market and random eigenpairs has been filtered out) for link occupation 
  values 1) $p=0.01$, 2) $p=0.03$, 3) $p=0.05$ and 4) $p=0.07$. Nodes are
  denoted as in Fig.~\ref{fig:assetgs}.}
\label{fig:assetg_filts}
\end{center}      
\end{figure}

\begin{figure}
\begin{center}
\includegraphics[width=250pt]{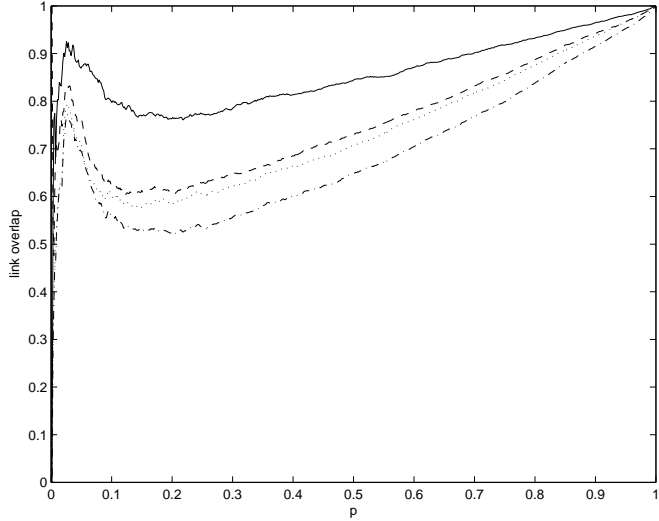}
\caption{The fraction of overlapping (i.e. common) links in asset graphs constructed
  from $\boldsymbol{C}_{g}$ and $\boldsymbol{C}_{-m}$ (solid line), $\boldsymbol{C}$ and $\boldsymbol{C}_{-m}$ (dashed
  line), $\boldsymbol{C}$ and $\boldsymbol{C}_{g}$ (dotted line) and in all three (dashdotted line).}
\label{fig:overlap}
\end{center}      
\end{figure}

In Figure
\ref{fig:assetg_filts} we show asset graphs constructed using $\boldsymbol{C}_{g}$ for link 
occupation values of $p=0.01$, $p=0.03$, $p=0.05$, and $p=0.07$. We see that 
these graphs are very similar to those presented in Fig.~\ref{fig:assetg_filts115}, 
\emph{i.e.}, to the ones constructed by using $\boldsymbol{C}_{-m}$. This is verified 
in Fig.~\ref{fig:overlap}, which shows the fraction of overlapping links, 
i.e. the percentage of common links, in the studied asset graphs. The
shape of the curves turns out to be interesting. In all cases the overlap 
increases very rapidly until $p \approx 0.025$. After this, the overlap
decreases indicating that the links become more random, but as the
number of links grows larger and the fraction of ``free'' places
decreases, the overlap starts to increase again.
As a reference, one can use Erdös-Rényi ensemble $G(m,N)$,
which consists of graphs of $N$ nodes and exactly $m$ links, such that 
each possible graph appearing with equal probability. The overlap for
$G(m,N)$ is clearly $p^{2}$.

The overlap between asset graphs constructed from $\boldsymbol{C}_{g}$ and $\boldsymbol{C}_{-m}$ is
found to be around 93\% for $p \approx 0.025$ and over 90\% in the interval
$[0.022,0.037]$. From panel a of Fig.~\ref{fig:number_p}, in which we show the number
of isolated components as a function of $p$, one sees that this
number is also at its highest in the interval,
meaning that these are the most relevant values of $p$ when studying
the clustering. This means that we do not gain much by
filtering out the random eigenpairs before constructing the asset
graphs. At the same time we may lose some significant information about 
the small clusters stored in the lowest ranking eigenpairs, as discussed 
in the previous section. One should notice that
cluster identification is more difficult when the full
correlation matrix is used, but the difference is not very large. 
When this is
combined with the fact that most information about the financial and
conglomerate companies is lost when the market eigenvector is filtered
out, it is evident that best results are obtained by using both $\boldsymbol{C}$ 
and $\boldsymbol{C}_{-m}$.

Kim \emph{et al.}~\cite{Jeong} have suggested that asset graphs constructed
from $\boldsymbol{C}_{g}$ have a well-defined critical threshold $p_{c}$, where many
isolated components merge into one giant component. They
construct the asset graphs by including all the links that correspond to a
correlation coefficient above a predetermined value $q$ and plot the
number of isolated components as a function of $q$. A similar plot for
the present data set is shown in panel b of Fig.~\ref{fig:number_p}. At 
the first sight it seems that there exists a clear critical threshold $q_{c}
\approx 0.1$ (seen as a sudden jump in the number of components) for
the asset graphs constructed from $\boldsymbol{C}_{g}$ but no
clear threshold for those constructed from the full correlation
matrix. However, it is perhaps a little misleading to speak about a critical 
threshold, since the one "seen" in panel b of Fig.~\ref{fig:number_p} is 
due to the fact that the elements of $\boldsymbol{C}_{g}$ are not uniformly distributed. 
From panel a of Fig.~\ref{fig:number_p} one sees that there is no clear 
threshold in none of these cases (as no sudden jumps are seen). 

To summarize, it seems that the noise present in the time series does not
change the cluster structure of the asset graphs, which is not very
surprising since only links corresponding to
the highest correlation coefficients are included. It also seems that 
there is no critical threshold $p_{c}$ in any of the studied cases. 
Therefore, useful information may be lost, while no benefit is gained,
if the random eigenpairs are filtered out before constructing the
asset graphs.

\section{Summary}

The aim of the present work was to investigate in complex systems the relationship between 
the spectral properties of correlation based matrices and the cluster
structure of the related networks. The network was constructed as a
complete graph where the weights were identified with elements taken
from the correlation matrix. We have chosen to study stock market data
since large amount of information has already been accumulated about them
and their spectral properties have also been studied in detail
\cite{RMT_Potters, RMT_Stanley}. Two data sets from the NYSE were
analyzed, one with lesser stocks appropriate for visualization and a
larger one with better statistical properties.

We started our study by analyzing the eigenvector corresponding to the largest
eigenvalue of the weight matrix and found to a very good first approximation
that the eigenvector components correspond to the strengths of the nodes
(i.e. companies). There is a systematic second order correction
roughly proportional to the nodal strengths, which is probably due to
the modular structure of the network.

The identification of the clusters using the high ranking eigenvectors
turned out to be a too formidable task. Therefore, we have chosen a
different path: Using independent information, we have given
interpretations to typical eigenvectors. Our results show that
there are eigenvectors which are well localized to a few industrial
branches. Surprisingly, such eigenvectors are not have always high ranking
i.e. correspond to a large eigenvalue. On the 
other hand, some high rank eigenvectors represent so many branches
that they are hardly distinguishable from the random case. Therefore,
we think that the eigenvectors are not appropriate in identifying the
modules of such networks. By using the diffusion matrix we had to arrive
to a similar conclusion, though it should be emphasized that there is
a strong overlap between the highest ranking eigenvectors of the weight
and diffusion matrices. 

Since direct network methods are known to be efficient in identifying
the hierarchical structure of correlation based networks
\cite{mantegna1, onnela:dyn, onnela:clust, tumminello}, we have
studied how the spectral methods can be combined with the asset graph
method based on thresholding. We have compared the asset graphs as
obtained from the noisy and denoised correlation matrices, where
denoising was carried out by using spectral information
\cite{Jeong}. It turned out that denoising has little effect on the
clusters of asset graphs. This is because of the hierarchical
structure of the clusters and due to the fact that thresholding picks
the high correlations where the noise is expected to play a
subordinate role.  Surprisingly enough similar denoising methods seem to
work efficiently when applied directly to portfolio optimization
\cite{Pafka}.

We can conclude that the identification of
clusters or communities is even more difficult in the case of highly connected
weighted networks. Spectral methods may lead to an overall description
of the properties of complex systems but they do not seem to be
appropriate for the classification problem without additional
information about the nodes of the related network.

\section{Acknowledgements}
TH, JS and KK are supported by the Academy of Finland (The Finnish Center of
Excellence program 2006-2011). JK and GT were partially supported by OTKA K60456.

\end{document}